\documentclass[%
 aip,amsmath,amssymb,reprint,jcp
]{revtex4-2}

\usepackage{graphicx}
\usepackage{dcolumn}
\usepackage{bm}

\usepackage[utf8]{inputenc}
\usepackage[T1]{fontenc}
\usepackage{mathptmx}
\usepackage{etoolbox}
\usepackage[dvipsnames]{xcolor}
\usepackage{braket}
\usepackage{siunitx}
\usepackage{booktabs}
\usepackage{hyperref}
\hypersetup{
    colorlinks=true,
    allcolors=RoyalBlue
}


\newcommand{\sqiswap}[1]{$\sqrt{i\mathrm{SWAP}}_{#1}$}

\makeatother
\begin{document}

\title[Multi-species MCTDHF for gate simulations in multi-qubit hardware]{Multi-species MCTDHF for gate simulations in multi-qubit hardware}

\author{Oskar Leinonen}
\email{oskar.leinonen@fys.uio.no}
\affiliation{ 
Department of Physics, University of Oslo, Oslo N-0316, Norway
}

\date{\today}

\begin{abstract}
Gate fidelities of quantum hardware are usually estimated with few-level models, in which the level structure and the couplings between the qubits are fixed in advance. Here, we apply the multi-species multiconfiguration time-dependent Hartree--Fock (MCTDHF) method to gate calculations at the hardware level, where the spectrum and the couplings instead follow from the dynamics of the system. Each qubit is treated as a separate species with its own time-dependent orbitals. As a test case, we engineer $\sqrt{i\mathrm{SWAP}}$ gates in a one-dimensional register of four anharmonic wells. First, we simulate the system with one electron per well and validate the method against an untruncated propagation of the register. We also construct a gate between the two outermost qubits from circuits of nearest-neighbor gates. The fidelity of these circuits is not reliably predicted by the individual gates, so the circuits have to be propagated as a whole. We then simulate the system with three spin-polarized electrons per well, where the gates fail through leakage out of the state $|11\rangle$. A calculation with too few orbitals misses this leakage and reports a high-fidelity gate instead. Capturing the leakage requires a large number of orbitals, which a multilayer form of the ansatz makes feasible. These results show that the method is a suitable tool for designing gates in multi-qubit hardware, where the errors of a gate are not confined to the computational subspace of the qubits it acts on.
\end{abstract}

\maketitle

\section{Introduction}
\label{sec:intro}
Over the past three decades, the multiconfiguration time-dependent Hartree (MCTDH) method~\cite{meyerMulticonfigurationalTimedependentHartree1990,mantheWavepacketDynamicsMulticonfiguration1992,beckMulticonfigurationTimedependentHartree2000, meyer2009multidimensional} has been the workhorse for solving the time-dependent Schrödinger equation in systems with many coupled degrees of freedom. In MCTDH, the wavefunction is expanded in a small set of time-dependent orbitals, each represented in a large time-independent primitive basis. Both the expansion coefficients and the orbitals are propagated according to equations of motion derived from the Dirac–Frenkel variational principle~\cite{hochstuhlTimedependentMulticonfigurationMethods2014, beckMulticonfigurationTimedependentHartree2000}. The orbitals therefore adapt to the dynamics as they unfold, significantly reducing the number of orbitals required for a given accuracy compared to a standard fixed basis. This has made numerically exact wavefunction dynamics routine for molecular systems with tens of coupled vibrational modes~\cite{meyerStudyingMolecularQuantum2012, raabMolecularDynamicsPyrazine1999}. The multilayer formulation (ML-MCTDH), which recursively expands each orbital in its own layer of time-dependent functions, extends the same approach to hundreds or thousands of degrees of freedom~\cite{wangMultilayerFormulationMulticonfiguration2003, mantheMultilayerMulticonfigurationalTimedependent2008,meyerStudyingMolecularQuantum2012}.

The core idea of MCTDH has also been applied to systems of indistinguishable particles. For fermions, antisymmetrization of the configurations leads to the multiconfiguration time-dependent Hartree--Fock (MCTDHF)~\cite{zanghelliniTestingMulticonfigurationTimedependent2004,caillatCorrelatedMultielectronSystems2005,hochstuhlTimedependentMulticonfigurationMethods2014} method, while symmetrization yields the MCTDH for Bosons (MCTDHB) method~\cite{streltsovRoleExcitedStates2007,alonMulticonfigurationalTimedependentHartree2008}. The two are treated on a common footing in the MCTDH-X family~\cite{lodeColloquiumMulticonfigurationalTimedependent2020, alonUnifiedViewMulticonfigurational2007}. The ansatz extends further to mixtures of several species, where each species carries its own set of orbitals and the (anti)symmetrization is restricted to within a species~\cite{alonMulticonfigurationalTimedependentHartree2007, alonRecursiveFormulationMulticonfigurational2012, caoUnifiedInitioApproach2017}. MCTDHF is an established tool for strong-field and attosecond electron dynamics~\cite{caillatCorrelatedMultielectronSystems2005}, MCTDHB and MCTDH-X for correlated ultracold gases in traps and optical lattices~\cite{lodeColloquiumMulticonfigurationalTimedependent2020}, and the multi-species formulations for atomic mixtures and for impurities immersed in them~\cite{kronkeNonequilibriumQuantumDynamics2013,mistakidisDissipativeCorrelatedDynamics2019}.

The MCTDH family of methods has also been applied outside chemistry. Some of that work
points toward quantum technology. Quantum circuits are commonly simulated with tensor networks~\cite{markovSimulatingQuantumComputation2008, berezutskiiTensorNetworksQuantum2025}. These methods are closely related to MCTDH: in tensor-network language, the ML-MCTDH wavefunction is a tree tensor network~\cite{larssonTensorNetworkView2024}, and the method has been used to simulate circuit-based quantum computers~\cite{ellerbrockMultilayerMulticonfigurationalApproach2020}, where the qubits are abstract two-level systems and the gates are applied as unitary matrices. 

At the hardware level of a quantum device, where a qubit is a two-level
subspace of a much larger spectrum, MCTDH has seen almost no use. The closest is a CNOT gate realized in the vibrational states of ammonia, where MCTDH was combined with optimal control theory to shape the driving pulse of the gate~\cite{schroderRealizationCNOTQuantum2009}, though in a molecule rather than a device. The properties that make MCTDH(F) effective in chemical dynamics are, however, the same properties a gate calculation for a device requires. Because the orbitals are time-dependent, they adapt to the instantaneous confining potential and drive. Leakage out of the qubit subspace and the shifts of the levels under the drive therefore emerge from the dynamics. The reduced few-level models used for most gate-fidelity estimates instead fix the level structure in advance. The anharmonicity is supplied as a parameter, and population can only leak through the channels written into the model, such as the $\ket{11}$--$\ket{02}$ crossing~\cite{ghoshHighfidelityControlledsigmaZGate2013, martinisFastAdiabaticQubit2014, willschGateerrorAnalysisSimulations2017}. In MCTDH(F), the particles are represented on a spatial grid and their two-body interaction is retained in full, so the qubit--qubit coupling follows from the geometry of the device. Few-level models instead take the coupling as a fixed number, whether a capacitive or exchange coupling, a blockade shift, or a Lamb--Dicke factor~\cite{krantzQuantumEngineersGuide2019, blaisCircuitQuantumElectrodynamics2021, burkardSemiconductorSpinQubits2023, haffnerQuantumComputingTrapped2008, saffmanQuantumInformationRydberg2010}, which is computed or measured in advance and does not change while the gate runs.

Furthermore, the adaptivity of the MCTDH(F) orbitals keeps the computational cost down, since far fewer orbitals are needed than in a fixed basis of comparable accuracy. So far, device-level calculations that resolve the spectrum have paid that cost. Such calculations include gates in superconducting circuits computed from the quantized circuit Hamiltonian~\cite{strauchQuantumLogicGates2003, larssenWhichSuperconductingQubit2026,dingHighFidelityFrequencyFlexibleTwoQubit2023,weissFastHighFidelityGates2022}, trapped-ion gates with the motional modes kept explicit~\cite{sorensenQuantumComputationIons1999}, and configuration-interaction calculations of gates in semiconductor double quantum dots~\cite{nielsenImplicationsSimultaneousRequirements2010, nielsenConfigurationInteractionCalculations2012, nowakTimedependentConfigurationinteractionSimulations2010,lesnickiSpinSeparationExchange2017} and of electrons on helium~\cite{beysengulovCoulombInteractionDrivenEntanglement2024, leinonen2026}. Where the qubit is a single collective coordinate the cost is mild and several qubits are within reach, but where it is built from interacting particles these calculations stop at one or two electrons.

In this work, we apply multi-species MCTDHF to gate calculations at the hardware level. As a test case, we use a model register of four anharmonic wells in one dimension, with parameters taken from an electrons-on-helium platform. Each well forms a qubit and is treated as a separate species with its own time-dependent orbitals. State averaging allows all 16 qubit states of the four-well register to be propagated simultaneously, and a multilayer form of the ansatz keeps the number of coefficients small when each well contains several electrons. We demonstrate the framework by engineering \sqiswap{} gates between nearest-neighbor qubits and between the two outermost qubits. In the latter case, we realize the gate either by a direct interaction or through circuits of nearest-neighbor gates.

We first place a single electron in each well, in which case the method reduces to MCTDH. Four orbitals per well then reproduce an untruncated propagation of the 16 qubit states to within $5\times10^{-4}$ in average gate fidelity. Since the number of
orbitals can be chosen separately for each well, the gates can be optimized at
reduced cost with a single orbital in the wells a gate does not act on. For the nearest-neighbor gates, this truncation locates the same optimum as a calculation with four orbitals per well, whereas for the direct gate between the outermost qubits it overestimates the fidelity and misses leakage that
makes the fidelity vary irregularly with the gate duration. For the circuits, reduced models in which the circuit is a product of gate matrices projected onto the qubit subspace provide neither an upper nor a
lower bound on the fidelity. A reliable prediction therefore requires propagating the full register through the complete circuit, which the compact MCTDH representation makes affordable. We then place three spin-polarized electrons in each well, which requires antisymmetrization within each well and the multilayer ansatz. In this case, the gates fail through leakage out of the $\ket{11}$ state, which a calculation with too few orbitals misses, reporting a high-fidelity gate instead.

The paper is organized as follows: Section~\ref{sec:mctdhf} presents the multi-species MCTDHF framework, including state averaging and the multilayer ansatz. Section~\ref{sec:application} introduces the four-well register, the gate engineering protocol and the routed long-range gates used as a test case. The results are presented in Sec.~\ref{sec:results}, first with one electron per well (Sec.~\ref{sec:fourwell}) and then with three (Sec.~\ref{sec:threewell}). Conclusions are given in Sec.~\ref{sec:conclusions}. Technical details of the method and the reference propagation are given in the appendices, and details of the model and the gate protocol in the supplementary material.

\section{Multi-species MCTDHF}
\label{sec:mctdhf}
In this section we describe the multi-species MCTDHF framework used in this work. The ansatz combines the MCTDHF treatment of fermions~\cite{hochstuhlTimedependentMulticonfigurationMethods2014} with the MCTDH formulations for mixtures, so that every species carries its own set of orbitals and antisymmetrization acts only within a species~\cite{alonMulticonfigurationalTimedependentHartree2007, alonRecursiveFormulationMulticonfigurational2012, kronkeNonequilibriumQuantumDynamics2013, caoUnifiedInitioApproach2017}. The wavefunction ansatz is given in Sec.~\ref{sec:ansatz}, the equations of motion in Sec.~\ref{sec:eom}, the primitive-basis representation in Sec.~\ref{sec:discretization}, and state averaging in Sec.~\ref{sec:stateavg}. We then extend the framework to a multilayer form, with the ansatz in Sec.~\ref{sec:ml} and its equations of motion in Sec.~\ref{sec:ml-eom}.

\subsection{Wavefunction ansatz}
\label{sec:ansatz}

The MCTDHF method~\cite{hochstuhlTimedependentMulticonfigurationMethods2014} expands the
wavefunction in a basis of Slater determinants built from $M$ orbitals that
evolve in time. The number $M$ sets the position between two familiar limits.
When the expansion holds a single determinant it reduces to the time-dependent
Hartree--Fock (TDHF) mean-field approximation, and when $M$ equals the size
$N_b$ of the underlying primitive basis it becomes the time-dependent full
configuration interaction (TD-FCI) method. The latter is exact but grows
exponentially with the number of electrons. Because the orbital expansion
coefficients evolve under equations of motion derived from the time-dependent
variational principle, the essential physics is captured with a much smaller
number of orbitals than a static basis would require, and a quite low value of
$M$ is therefore often
sufficient~\cite{hochstuhlTimedependentMulticonfigurationMethods2014,
beckMulticonfigurationTimedependentHartree2000}.

In standard MCTDHF calculations of electronic systems whose Hamiltonian
conserves $S_z$, the spin-unrestricted formulation treats the $\alpha$- and
$\beta$-spin electrons as distinguishable. Each spin is expanded in its own set of time-dependent orbitals, with its own orbital equation of motion~\cite{hochstuhlTimedependentMulticonfigurationMethods2014}. This split is exact,
since $S_z$ conservation forbids the spins from mixing, and it is more
compact, since each orbital set only needs to capture the correlation within its
own spin. The same construction extends from two spin species to an arbitrary number of distinguishable species, as in the MCTDH formulations for mixtures~\cite{alonMulticonfigurationalTimedependentHartree2007, alonRecursiveFormulationMulticonfigurational2012, caoUnifiedInitioApproach2017}. In simulations of quantum computing hardware, each species can represent a different subsystem, such as a qubit. We introduce a
species label $\sigma = 0,\dots,K-1$ for $K$ species, with $N_\sigma$ particles
in species $\sigma$. The wavefunction is then expanded as

\begin{equation}
  |\Psi(t)\rangle = \sum_{I} C_I(t)\,|I(t)\rangle ,
  \qquad
  |I(t)\rangle = \bigotimes_{\sigma}|I_\sigma(t)\rangle ,
  \label{eq:ansatz}
\end{equation}
where $|I_\sigma(t)\rangle$ is a Slater determinant of $N_\sigma$
particles drawn from the $M_\sigma$ time-dependent species orbitals
$\{|\phi^\sigma_n(t)\rangle\}_{n=1}^{M_\sigma}$, and the composite index $I=(I_0,\dots,I_{K-1})$
collects the determinant indices $I_\sigma$ of all species. Each species orbital is in turn represented in a time-independent primitive
basis $\{|\chi^{\sigma}_k\rangle\}$ of size $N_{b,\sigma}$,
\begin{equation}
  |\phi^{\sigma}_n(t)\rangle
  = \sum_{k=1}^{N_{b,\sigma}} b^{\sigma}_{kn}(t)\,|\chi^{\sigma}_k\rangle.
  \label{eq:bexp}
\end{equation}
The coefficients $b^{\sigma}_{kn}(t) = \langle
\chi^{\sigma}_k|\phi^{\sigma}_n(t)\rangle$ are the orbital variational
parameters. Together with the configuration coefficients $C_I(t)$ of
Eq.~\eqref{eq:ansatz}, they form the complete set of variational parameters
of the ansatz. 

\subsection{Equations of motion}
\label{sec:eom}
The equations of motion follow from applying the Dirac--Frenkel variational principle to Eq.~\eqref{eq:ansatz}. Their input is the Hamiltonian, which in second quantization reads
\begin{equation}
\begin{split}
  \hat H(t) &= \sum_{pq,\sigma} h^{\sigma}_{pq}(t)\,
              \hat a^{\dagger}_{p\sigma}\hat a_{q\sigma} \\
            &\quad + \frac{1}{2}\sum_{pqrs,\sigma\tau} g^{\sigma\tau}_{pqrs}(t)\,
              \hat a^{\dagger}_{p\sigma}\hat a^{\dagger}_{r\tau}
              \hat a_{s\tau}\hat a_{q\sigma},
\end{split}
  \label{eq:ham}
\end{equation}
where $h^{\sigma}_{pq}(t)$ and $g^{\sigma\tau}_{pqrs}(t)$ are the one- and two-body matrix elements in the species orbitals. They are obtained by transforming the matrix elements in the primitive basis of Eq.~\eqref{eq:bexp} (Sec.~\ref{sec:discretization}) and therefore depend on time through the orbitals. In this work only the one-body part also carries explicit time dependence, so the primitive-basis two-body elements are constant.
The diagonal blocks $\sigma = \tau$ of $g^{\sigma\tau}_{pqrs}$ are the intra-species interaction and the off-diagonal blocks $\sigma \neq \tau$ the inter-species one.

We vary the coefficients and the orbitals independently, keeping the orbitals of each species orthonormal. Rotations among the orbitals of one species leave the state unchanged, since they can be absorbed into the coefficients. We fix this freedom by requiring $\langle\phi^{\sigma}_p|\dot\phi^{\sigma}_q\rangle = 0$, so that the orbitals move only out of their current span. Following the derivation of Ref.~\cite{hochstuhlTimedependentMulticonfigurationMethods2014} with the spin sum replaced by a sum over species, as for mixtures of identical particles~\cite{alonRecursiveFormulationMulticonfigurational2012}, we obtain coupled equations of motion for the coefficients and the orbitals. The coefficients evolve according to
\begin{equation}
  i\,\frac{\partial}{\partial t} C_I(t)
  = \sum_J \langle I|\hat H(t)|J\rangle\, C_J(t),
  \label{eq:ci}
\end{equation}
and the orbitals of each species satisfy
\begin{equation}
\begin{split}
  i\,\frac{\partial}{\partial t}|\phi^{\sigma}_n\rangle
  = \hat P^{\sigma}\Bigl[ &\,
      \hat h^{\sigma}(t)|\phi^{\sigma}_n\rangle \\
    &+ \sum_{pqrs,\tau}
      \bigl[(D^{\sigma})^{-1}\bigr]_{np}\,
      d^{\sigma\tau}_{pqrs}\,
      \hat g^{\sigma\tau}_{rs}\,|\phi^{\sigma}_q\rangle
    \Bigr],
\end{split}
  \label{eq:orb}
\end{equation}
where $D^{\sigma}_{pq} = \langle\Psi|\hat a^{\dagger}_{p\sigma}\hat
a_{q\sigma}|\Psi\rangle$ and $d^{\sigma\tau}_{pqrs} = \langle\Psi|\hat
a^{\dagger}_{p\sigma}\hat a^{\dagger}_{r\tau}\hat a_{s\tau}\hat
a_{q\sigma}|\Psi\rangle$ are the one- and two-particle reduced density
matrices, $\hat g^{\sigma\tau}_{rs}$ is the mean-field operator formed from the
orbitals $\phi^{\tau}_r$ and $\phi^{\tau}_s$ of species $\tau$, and $\hat
P^{\sigma} = 1 - \sum_m |\phi^{\sigma}_m\rangle\langle\phi^{\sigma}_m|$
projects onto the complement of the orbital space of species $\sigma$.

The multi-species content of Eq.~\eqref{eq:orb} is threefold. The partner sum over $\tau$ runs over all $K$ species, so each species propagates in the instantaneous mean field of every other, and the species couple only through $d^{\sigma\tau}_{pqrs}$ and $\hat g^{\sigma\tau}_{rs}$. Each species carries its own projector and density matrix, so the orbital equations are structurally independent and may be integrated on distinct primitive grids. The inversion of $D^{\sigma}$ is likewise per species, so the conditioning is set only by the natural occupations \emph{within} a species.

Because the orbitals move, the Hamiltonian matrix in Eq.~\eqref{eq:ci} changes at every step and is never formed. Its action is instead evaluated directly~\cite{hochstuhlTimedependentMulticonfigurationMethods2014} as a sum over species and species pairs (App.~\ref{app:eom}).

\subsection{Primitive-basis representation}
\label{sec:discretization}

Throughout this work the primitive basis $\{|\chi^{\sigma}_k\rangle\}$
of Eq.~\eqref{eq:bexp} is a sine-DVR (discrete variable representation)~\cite{beckMulticonfigurationTimedependentHartree2000, colbertNovelDiscreteVariable1992}
basis on each species' spatial grid, orthonormal by construction. The sine-DVR basis allows both the one- and two-body matrix elements to be evaluated directly on the grid. Using the orthonormality of the basis, projecting Eq.~\eqref{eq:orb} onto $\langle\chi^{\sigma}_k|$ and acting with the projection operator $\hat P^{\sigma}$ gives the discretized working form of the orbital equation of motion,

\begin{equation}
\begin{split}
  i\,\dot b^{\sigma}_{kn}
  &= h^{\sigma,(1)}_{kn} - h^{\sigma,(3)}_{kn} \\
  &\quad + \sum_{pqrs,\tau}
    \bigl[(D^{\sigma})^{-1}\bigr]_{np}\,
    d^{\sigma\tau}_{pqrs}
    \left( g^{\sigma\tau,(3)}_{kqrs} - g^{\sigma\tau,(5)}_{kqrs} \right).
\end{split}
  \label{eq:bdot}
\end{equation}
The superscripts, following
Ref.~\cite{hochstuhlTimedependentMulticonfigurationMethods2014}, record the
transformation level between the primitive and species-orbital bases.
$h^{\sigma,(1)}_{kn} = \langle\chi^{\sigma}_k|\hat h^{\sigma}(t)|\phi^{\sigma}_n
\rangle$ retains one primitive index, while $h^{\sigma,(3)}_{kn} =
\langle\chi^{\sigma}_k|(1-\hat P^{\sigma})\hat h^{\sigma}(t)|\phi^{\sigma}_n
\rangle$ carries an additional projection onto the orbitals of species
$\sigma$. The two-body matrix elements $g^{\sigma\tau,(3)}_{kqrs}$ and $g^{\sigma\tau,(5)}_{kqrs}$ are
defined the same way with $\hat g^{\sigma\tau}_{rs}$ in place of $\hat
h^{\sigma}$. Each difference in Eq.~\eqref{eq:bdot} is thus the
corresponding term of Eq.~\eqref{eq:orb}, projected onto the primitive
basis:
$h^{\sigma,(1)}_{kn}-h^{\sigma,(3)}_{kn}=\langle\chi^{\sigma}_k|\hat
P^{\sigma}\hat h^{\sigma}(t)|\phi^{\sigma}_n\rangle$ and
$g^{\sigma\tau,(3)}_{kqrs}-g^{\sigma\tau,(5)}_{kqrs}=\langle\chi^{\sigma}_k|\hat
P^{\sigma}\hat g^{\sigma\tau}_{rs}|\phi^{\sigma}_q\rangle$.
The fully transformed elements, $h^{\sigma,(2)}_{pq}=h^{\sigma}_{pq}$ and
$g^{\sigma\tau,(4)}_{pqrs}=g^{\sigma\tau}_{pqrs}$, are those of Eq.~\eqref{eq:ham}
and enter the coefficient equation of motion, Eq.~\eqref{eq:ci}.

\subsection{State averaging}
\label{sec:stateavg}

The equations of motion above describe the propagation of a single wavefunction $|\Psi(t)\rangle$. In order to propagate multiple states simultaneously, we use a state-averaged MCTDHF approach, in which the orbitals are optimized for a weighted average of the states instead of for a single one~\cite{wernerQuadraticallyConvergentMCSCF1981}. Each state $|\Psi_N(t)\rangle$ has its own set of coefficients $C_I^N(t)$, which evolve according to Eq.~\eqref{eq:ci}, while the orbitals are shared between all propagated states. The density matrices entering Eq.~\eqref{eq:orb} are thus replaced by
\begin{equation}
\begin{split}
  D^{\sigma}_{pq} &\to \sum_N w_N\,
    \langle\Psi_N|\hat a^{\dagger}_{p\sigma}\hat a_{q\sigma}|\Psi_N\rangle, \\
  d^{\sigma\tau}_{pqrs} &\to \sum_N w_N\,
    \langle\Psi_N|\hat a^{\dagger}_{p\sigma}\hat a^{\dagger}_{r\tau}
    \hat a_{s\tau}\hat a_{q\sigma}|\Psi_N\rangle,
\end{split}
  \label{eq:densities-sa}
\end{equation}
with weights $w_N$ summing to one. The weights are free parameters and can be chosen based on which of the propagated states are of primary interest. With this replacement, the orbital equation of motion, Eq.~\eqref{eq:orb}, and its discretized form, Eq.~\eqref{eq:bdot}, are otherwise unchanged.

\subsection{Multilayer ansatz}
\label{sec:ml}
\begin{figure}[ht]
  \centering
  \includegraphics{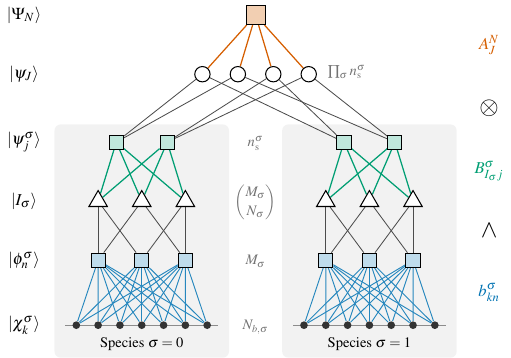}
  \caption{Multilayer MCTDHF ansatz, illustrated for $K=2$ species with
$N_\sigma=2$ particles in $M_\sigma=3$ orbitals and $n^{\sigma}_{\mathrm s}=2$
SBS per species. Nodes are functions and edges show how each layer is built
from the one below. Squares are linear combinations, with the variational
coefficients on the colored edges: the orbitals $\phi^\sigma_n$ are expanded in
the sine-DVR functions $\chi^\sigma_k$ through $b^\sigma$ (Eq.~\eqref{eq:bexp}), the SBS
$\psi^\sigma_j$ in the determinants $I_\sigma$ through $B^\sigma$
(Eq.~\eqref{eq:sbs}), and the state $\Psi_N$ in the product states $\psi_J$
through $A^N$ (Eq.~\eqref{eq:ml-ansatz}). The determinants (triangles) are
antisymmetrized products ($\wedge$) of $N_\sigma$ orbitals of one species, and
the product states (circles) are tensor products ($\otimes$) of one SBS per
species. The number of functions in each layer is
shown in gray.}
\label{fig:ml}
\end{figure}

The cost of the ansatz in Eq.~\eqref{eq:ansatz} is set by the number of configuration
coefficients $C_I$. For each state there are $\prod_\sigma N^{\mathrm{det}}_\sigma$ of them, where $N^{\mathrm{det}}_\sigma=\binom{M_\sigma}{N_\sigma}$ is the number
of determinants $|I_\sigma\rangle$ that can be built from the $M_\sigma$ orbitals of species $\sigma$. For three electrons per well, $N^{\mathrm{det}}_\sigma$ grows as $M_\sigma^3$, so the number of coefficients for four wells grows as $M_\sigma^{12}$ and quickly becomes prohibitively large to store. To reduce the cost, we use a multilayer
ansatz~\cite{wangMultilayerFormulationMulticonfiguration2003, mantheMultilayerMulticonfigurationalTimedependent2008, vendrellMultilayerMulticonfigurationTimedependent2011, kronkeNonequilibriumQuantumDynamics2013, caoUnifiedInitioApproach2017},
illustrated in Fig.~\ref{fig:ml}. In this ansatz, the determinants of each species are contracted into a smaller set of species
basis states (SBS), each a linear combination of the determinants $|I_\sigma\rangle$,
\begin{equation}
  |\psi^{\sigma}_j(t)\rangle = \sum_{I_\sigma=1}^{N^{\mathrm{det}}_\sigma} B^{\sigma}_{I_\sigma j}(t)\,|I_\sigma(t)\rangle , \quad j=1,\dots,n^{\sigma}_{\mathrm s},
  \label{eq:sbs}
\end{equation}
where $n^{\sigma}_{\mathrm s}\le N^{\mathrm{det}}_\sigma$ is the number of SBS of
species $\sigma$. The expansion matrices
$B^\sigma$ have orthonormal columns, $(B^{\sigma})^\dagger B^{\sigma}=1$. For
$n^{\sigma}_{\mathrm s}=N^{\mathrm{det}}_\sigma$ they are unitary, so the SBS span
the same space as the determinants. Each SBS is a function of all $N_\sigma$ particles of species $\sigma$, whereas the orbitals $\phi^\sigma_n$ of Eq.~\eqref{eq:bexp} are functions of one particle.

Each propagated state $|\Psi_N\rangle$ of Sec.~\ref{sec:stateavg} is expanded in the
SBS, similar to the expansion in determinants of the single-layer ansatz in Eq.~\eqref{eq:ansatz},
\begin{equation}
  |\Psi_N(t)\rangle = \sum_{J} A^{N}_{J}(t)\,|\psi_{J}(t)\rangle ,
  \qquad
  |\psi_{J}(t)\rangle = \bigotimes_{\sigma}|\psi^{\sigma}_{j_\sigma}(t)\rangle ,
  \label{eq:ml-ansatz}
\end{equation}
where the composite index $J=(j_0,\dots,j_{K-1})$ collects the SBS indices $j_\sigma$ of all species.
The SBS and the orbitals are shared by all states, and only the top tensor $A^N_J$
depends on the state number $N$. For each state, the $\prod_\sigma N^{\mathrm{det}}_\sigma$ configuration
coefficients $C_I$ of the standard ansatz are now replaced by only
$\prod_\sigma n^{\sigma}_{\mathrm s}$ elements of $A^N_J$, a large reduction whenever
$n^{\sigma}_{\mathrm s}\ll N^{\mathrm{det}}_\sigma$. 

The multilayer ansatz has three layers of variational parameters: the top tensor $A^N$, the SBS coefficients $B^\sigma$ and
the orbital coefficients $b^\sigma$ of Eq.~\eqref{eq:bexp}. The ansatz is truncated by two parameters per species, the number of orbitals $M_\sigma$
and the number of SBS $n^{\sigma}_{\mathrm s}$, and is exact within the primitive basis for
$M_\sigma=N_{b,\sigma}$ and $n^{\sigma}_{\mathrm s}=N^{\mathrm{det}}_\sigma$.
The configuration coefficients are recovered as $C^N_{I}=\sum_{J}A^N_{J}\prod_\sigma
B^\sigma_{I_\sigma j_\sigma}$. For $N_\sigma=1$ the SBS are single-particle
functions that can be absorbed into the orbitals, so the middle layer is redundant.

\subsection{Multilayer equations of motion}
\label{sec:ml-eom}
The equations of motion follow from the Dirac--Frenkel variational principle, as in
Sec.~\ref{sec:eom}. The gauge freedom of each layer is fixed by
$\langle\phi^{\sigma}_p|\dot\phi^{\sigma}_q\rangle=0$ for the orbitals and
$(B^{\sigma})^\dagger\dot B^{\sigma}=0$ for the SBS coefficients. The top tensor
then obeys Eq.~\eqref{eq:ci} in the product basis $|\psi_J\rangle$,
\begin{equation}
  i\,\dot A^{N}_{J} = \sum_{J'}\langle\psi_{J}|\hat H|\psi_{J'}\rangle\,A^{N}_{J'} ,
  \label{eq:ml-A}
\end{equation}
and the SBS coefficients of species $\sigma$ satisfy
\begin{equation}
\begin{split}
  i\,\dot B^{\sigma}_{I_\sigma j} = \langle I_\sigma|\hat P^{\sigma}_{\psi}\Bigl[&\hat H^{\sigma}|\psi^{\sigma}_j\rangle \\
  &+ \sum_{lm}\bigl[(\rho^{\sigma})^{-1}\bigr]_{jl}\, \langle\hat W\rangle^{\sigma}_{lm}\,|\psi^{\sigma}_m\rangle\Bigr] .
\end{split}
  \label{eq:ml-B}
\end{equation}
Here $\hat H^{\sigma}$ contains the terms of Eq.~\eqref{eq:ham} that act on species $\sigma$ alone,
$\hat W^{\sigma}$ is the interaction between species $\sigma$ and all other species, and
$\hat P^{\sigma}_{\psi}=1-\sum_l|\psi^\sigma_l\rangle\langle\psi^\sigma_l|$ projects onto the
complement of the SBS space. The left-hand side is the change of the SBS due to $B^\sigma$
at fixed orbitals, while the change due to the motion of the orbitals is given by Eq.~\eqref{eq:orb}.
The density matrix $\rho^\sigma_{jl}$ and the mean-field operators $\langle\hat W\rangle^{\sigma}_{lm}$ of
species $\sigma$ are
\begin{equation}
\begin{split}
  \rho^{\sigma}_{jl} &= \sum_N w_N\,\langle\Psi^{\sigma}_{N,j}|\Psi^{\sigma}_{N,l}\rangle, \\
  \langle\hat W\rangle^{\sigma}_{lm} &= \sum_N w_N\,\langle\Psi^{\sigma}_{N,l}|\hat W^{\sigma}|\Psi^{\sigma}_{N,m}\rangle ,
\end{split}
  \label{eq:ml-rho}
\end{equation}
with the single-hole functions $|\Psi^{\sigma}_{N,j}\rangle=\langle\psi^\sigma_j|\Psi_N\rangle$ and the state-averaging weights $w_N$ of Sec.~\ref{sec:stateavg}. The intra-species term of Eq.~\eqref{eq:ml-B} carries no $(\rho^\sigma)^{-1}$, because the factor cancels against $\rho^\sigma$.

The orbitals evolve as in the single-layer case. Equations~\eqref{eq:orb} and~\eqref{eq:bdot} are unchanged, except that the density matrices $D^\sigma_{pq}$ and $d^{\sigma\tau}_{pqrs}$ are evaluated from $A^N$ and $B^\sigma$. The propagation scheme is described in App.~\ref{app:propagation}.

\section{Application to multi-qubit hardware}
\label{sec:application}
The multi-species MCTDH(F) framework is particularly well suited to simulating multi-qubit hardware, since qubits by construction represent distinguishable subsystems. We demonstrate this framework by engineering two-qubit gates in a model system of four anharmonic wells. In this section, we introduce the model system, describe the two-qubit gate engineering protocol, and how we have applied MCTDH(F) to the problem. 

\subsection{The four anharmonic well system}
\label{sec:sys-fourwell}
The system we consider is a one-dimensional register of four anharmonic wells. The barriers between the wells are high enough that tunneling between them is negligible. The four wells can therefore be treated as distinguishable subsystems, which we identify with the $K=4$ species $\sigma=0,\dots,3$ of Sec.~\ref{sec:mctdhf}. Well $\sigma$ holds $N_\sigma$ electrons, and we denote the position of its $i$th electron by $x^{\sigma}_i$. The one-body operator acting on a single electron of well $\sigma$ is
\begin{equation}
  \hat h^{\sigma}(x) =
-\frac{1}{2}\frac{\partial^2}{\partial x^2} + V^{\sigma}(x),
  \label{eq:h1}
\end{equation}
where $V^{\sigma}(x)$ is the anharmonic potential
\begin{equation}
  V^{\sigma}(x) = \frac{1}{2}(\omega^{\sigma})^2 (x-x_c^{\sigma})^2
  + \frac{\beta^{\sigma}(\omega^{\sigma})^2}{3}(x-x_c^{\sigma})^4.
  \label{eq:well}
\end{equation}
Here, $\omega^{\sigma}$, $\beta^{\sigma}$, and $x_c^{\sigma}$ denote the frequency, first-order anharmonicity (see Sec.~\ref*{S-app:anharm} of the supplementary material), and well center for subsystem $\sigma$. We set $\hbar=m_e=1$ throughout this work, and report $\omega^\sigma/2\pi$ and $\beta^\sigma/2\pi$ in GHz. The four well centers are fixed at $\SI{\pm1}{\micro\meter}$ and $\SI{\pm3}{\micro\meter}$. Each well is represented on its own sine-DVR grid, with 64 points for one electron per well and 80 points for three electrons per well (App.~\ref{app:grid}). The grids are \SI{1}{\micro\meter} wide, which makes them spatially disjoint. Consequently, electrons in different wells can interact through the bare Coulomb potential $\kappa/\lvert x-x'\rvert$, where $\kappa=e^2/4\pi\varepsilon_0$ is the interaction strength, without encountering singularities. For electrons in the same well, we derive an effective one-dimensional Coulomb interaction $V_{\mathrm{eff}}(x-x')$ by temporarily introducing transverse degrees of freedom and integrating them out (Sec.~\ref*{S-app:veff} of the supplementary material). To ensure physical realism, the wavefunction along the transverse coordinates, the interaction strength $\kappa$, and the well parameters ($\omega^{\sigma}, \beta^{\sigma}$) are all modeled after an electrons-on-helium platform~\cite{leinonen2026,beysengulovCoulombInteractionDrivenEntanglement2024}. The full Hamiltonian, acting on the coordinates of all $\sum_\sigma N_\sigma$ electrons, is then
\begin{equation}
\begin{split}
  \hat H
  &= \sum_{\sigma=0}^{3}\Bigl[\,\sum_{i=1}^{N_\sigma}\hat h^{\sigma}(x^{\sigma}_i)
     + \frac{1}{2}\sum_{i\neq j}^{N_\sigma} V_{\mathrm{eff}}\bigl(x^{\sigma}_i-x^{\sigma}_j\bigr)\Bigr] \\
  &\quad + \frac{1}{2}\sum_{\sigma\neq\tau}\sum_{i=1}^{N_\sigma}\sum_{j=1}^{N_\tau}
     \frac{\kappa}{\lvert x^{\sigma}_i-x^{\tau}_j\rvert}.
\end{split}
\label{eq:ham-fourwell}
\end{equation}
The first line contains all terms that involve a single well, i.e.\ the one-body terms and the intra-species interaction of Eq.~\eqref{eq:ham}, and the second line is the inter-species interaction. In the gate protocol of Sec.~\ref{sec:td-opt} the Hamiltonian becomes time-dependent through the well parameters, i.e.\ $\hat h^{\sigma} = \hat h^{\sigma}(t)$. We leave this implicit here.

We consider two cases for the number of electrons per well. In the first, each well holds a single electron,
$N_\sigma=1$. The intra-well interaction in Eq.~\eqref{eq:ham-fourwell} then
vanishes, and since each species consists of a single particle there is no
antisymmetry to impose. The multi-species MCTDHF ansatz therefore reduces to
MCTDH, and the middle layer of the multilayer ansatz is redundant. In the second case, each well holds three
spin-polarized electrons, $N_\sigma=3$. The intra-well interaction term in Eq.~\eqref{eq:ham-fourwell} is then present, and because the electrons within a well
share the same spin, their spatial wavefunction must be antisymmetric. This case therefore requires the full MCTDHF framework of Sec.~\ref{sec:mctdhf}.
Because the number of configurations grows rapidly with $N_\sigma$, we use its multilayer form
(Secs.~\ref{sec:ml} and~\ref{sec:ml-eom}). The gate protocol of Secs.~\ref{sec:op-points}--\ref{sec:fidelity} is applied to both cases. The long-range gate test of Sec.~\ref{sec:longrange-test} is applied only with one electron per well.

\subsection{Two-qubit gate engineering}
\label{sec:gate-engineering}
The two-qubit gate engineering protocol consists of two main steps. First, we optimize the static parameters of the time-independent Hamiltonian to identify operating points with target properties. Second, we drive the system between these operating points using a time-dependent Hamiltonian, and the functional form of the time dependence is optimized to achieve the desired gate operation. These two steps are described in detail below.

\subsubsection{Optimization of operating points}
\label{sec:op-points}
The properties of the four anharmonic well system can be controlled by tuning the well parameters $\vec{\lambda}=\{\omega^{\sigma},\beta^{\sigma}\}_{\sigma=0}^3$. We call a specific setting of these parameters an operating point. We propagate the state-averaged equations of motion in imaginary time to find the lowest-energy eigenstates of the Hamiltonian for a given operating point, using the MCTDHF ansatz of Sec.~\ref{sec:ansatz} with one electron per well and the multilayer ansatz of Sec.~\ref{sec:ml} with three electrons per well. Note that the orbitals and the CI coefficients retain a residual gauge freedom: a static joint rotation of the two leaves the physical states unchanged. We fix this gauge by diagonalizing the state-averaged one-particle density matrix, i.e.\ working in its natural-orbital basis. In the multilayer ansatz the corresponding freedom lies between the SBS and the top tensor, and we fix it in the natural-SBS basis, i.e.\ by diagonalizing the state-averaged SBS density matrix of Eq.~\eqref{eq:ml-rho}.
The gauge-fixed eigenstates are then used to evaluate a cost function that quantifies how well the operating point achieves the desired system properties. The system parameters are then optimized using a derivative-free optimizer. Additional details of the optimization are given in Sec.~\ref*{S-app:op-points} of the supplementary material. 

First, we optimize for an \emph{idle} operating point, where the qubits are decoupled. In the MCTDHF formalism this means that each eigenstate is dominated by a single configuration $\ket{I}$ of Eq.~\eqref{eq:ansatz}, with $\lvert C_I \rvert^2\approx1$. We write out the compound index $I$ in terms of the well occupations $(n_0, n_1, n_2, n_3)$ and use a weighted optimal assignment procedure to pair the qubit states $\ket{\vec n} = \ket{n_0n_1n_2n_3}$ with the eigenstates based on the coefficients $C_{n_0n_1n_2n_3}$. In other words, an ideal idle point has one eigenstate with $\lvert C_{0000} \rvert^2 = 1$, defining the qubit state $\ket{0000}$, one eigenstate with $\lvert C_{1000} \rvert^2 = 1$, defining the qubit state $\ket{1000}$, and so on for all 16 qubit states. In the optimization for the idle operating point, we require the well frequencies to be ascending, $\omega^0 < \omega^1 < \omega^2 < \omega^3$. With three electrons per well the same statements hold for the top-tensor coefficients $A_{n_0n_1n_2n_3}$ in the natural-SBS gauge, where $\ket{n_\sigma}_\sigma$ are the local basis states of well $\sigma$, as described in Sec.~\ref*{S-app:op-points-3e} of the supplementary material.

To achieve two-qubit gates, we also need \emph{target} operating points, where the qubits are coupled in a way that enables a gate operation. In this work we focus on \sqiswap{} gates. The two qubits a gate acts on, the \emph{active} qubits, are tuned so that two eigenstates become equal superpositions of the corresponding single-excitation qubit states, satisfying $\lvert C_I \rvert^2 = 1/2$. For example, for the gate between well 0 and well 1, two eigenstates should take the form $(|1000\rangle \pm |0100\rangle)/\sqrt{2}$, i.e.\ have coefficients $|C_{1000}|^2 = |C_{0100}|^2 = 1/2$. In the optimization for the target operating points, only the well parameters of the active wells are allowed to vary, while the wells of the two remaining qubits, the \emph{spectator} qubits, are held at their idle values.

In all of these optimizations, we restrict the frequencies $\omega^{\sigma}/2\pi$ to the range $[\num{5}, \num{15}]~\si{\giga\hertz}$ and the anharmonicities $\beta^{\sigma}/2\pi$ to $[\num{0.05}, \num{2}]~\si{\giga\hertz}$, which are physically reasonable for electrons on helium~\cite{beysengulovCoulombInteractionDrivenEntanglement2024}. 

\subsubsection{Optimization of the control schedule}
\label{sec:td-opt}
Once the operating points are fixed, the gate is realized by giving
the system parameters a time dependence $\vec{\lambda}(t)$ that drives the
system from the idle point to the target point and back. In this work, we use the following time-dependent function,

\begin{equation}
  \vec{\lambda}(t) = \vec{\lambda}_{\mathrm{idle}}
  + \frac{\vec{\lambda}_{\mathrm{target}}-\vec{\lambda}_{\mathrm{idle}}}{2}
  \bigl[ R(t) - R(t-t_{\mathrm{ramp}}-t_{\mathrm{hold}}) \bigr],
  \label{eq:time_dep_schedule}
\end{equation}
with
\begin{equation}
  R(t) = \mathrm{erf}\!\left(\frac{6}{t_{\mathrm{ramp}}}
         \Bigl(t-\frac{t_{\mathrm{ramp}}}{2}\Bigr)\right).
  \label{eq:ramp}
\end{equation}
This function smoothly interpolates from the idle operating point $\vec{\lambda}_{\mathrm{idle}}$ to the target operating point $\vec{\lambda}_{\mathrm{target}}$ over a ramp time $t_{\mathrm{ramp}}$, holds near $\vec{\lambda}_{\mathrm{target}}$ for a duration $t_{\mathrm{hold}}$, and ramps back down to idle, giving a total gate duration $t_{\mathrm{gate}} = 2t_{\mathrm{ramp}} + t_{\mathrm{hold}}$. 

With the static operating points $\vec{\lambda}_{\mathrm{idle}}$ and $\vec{\lambda}_{\mathrm{target}}$ fixed, the optimization of the control schedule relies entirely on tuning the timing parameters $t_{\mathrm{ramp}}$ and $t_{\mathrm{hold}}$. We optimize these parameters via a grid search over $(t_{\mathrm{ramp}}, t_{\mathrm{hold}})$, propagating the system in real time using the MCTDHF equations of motion and selecting the pair that maximizes the average gate fidelity $\bar F_4$ of the active pair (Sec.~\ref{sec:fidelity}). To reduce the cost of the grid search, we treat the two spectator wells at the mean-field level by restricting them to $N_\sigma$ orbitals each. This prevents the spectators from being excited and neglects the correlation between the active and spectator wells. To assess the accuracy of this approximation, we repeat the hold-time sweep at the optimal ramp time with the full orbital count in all wells. These results are presented in Sec.~\ref{sec:mf-res}. Numerical details of the real-time propagation are given in App.~\ref{app:propagation}.

\subsubsection{Gate fidelity and fidelity decomposition}
\label{sec:fidelity}
We quantify the quality of the gates by the average gate fidelity~\cite{nielsenSimpleFormulaAverage2002} 
\begin{equation}
      \bar F_d = \frac{\mathrm{Tr}(MM^{\dagger})
  + \max_{\vec{\theta},\vec{\phi}}\bigl\lvert\mathrm{Tr}\bigl(U^{\dagger} R_z^{\otimes n}(\vec{\theta})\, M\, R_z^{\otimes n}(\vec{\phi})\bigr)\bigr\rvert^2}
  {d(d+1)},
  \label{eq:fidelity}
\end{equation}
where $M_{ij}=\braket{\Phi_i|\Psi_j(t_{\mathrm{gate}})}$ and $d$ is the dimension of the qubit subspace being evaluated. Here $\ket{\Phi_i}$ is the eigenstate of the idle operating point that defines the $i$th qubit state, and $\ket{\Psi_j(t)}$ is $\ket{\Phi_j}$ propagated under the control schedule. The matrix $M$ is thus the transformation realized by the gate within the qubit subspace, with $M_{ij}$ the complex amplitude for qubit state $j$ to end up in qubit state $i$. The matrix $U$ is the ideal gate matrix, so that $M=U$ for a perfect gate. The maximization runs over independent $z$-rotations of each qubit on the output and input sides of $M$, $R_z^{\otimes n}(\vec{\theta})=\bigotimes_{k=1}^{n}R_z(\theta_k)$ with $R_z(\theta)=e^{-i\theta Z/2}$ and $\vec{\theta},\vec{\phi}\in[0,2\pi)^{n}$, where $n$ is the number of qubits. These two rotations remove single-qubit phases that carry no information about the entangling operation itself.

The infidelity, $1-\bar F_d$, can be decomposed into distinct error channels by exploiting the fact that Eq.~\eqref{eq:fidelity} depends on the propagated states solely through two quantities for each state $j$: the population $s_j$ retained within the qubit subspace, and the overlap $o_j$ of its realized output with the correct gate transformation $U\ket{\Phi_j}$,
\begin{equation}
  s_j=\sum_i \lvert M_{ij}\rvert^2, \qquad
    o_j=\sum_i U^*_{ij}\,\bigl(R_z^{\otimes n}(\vec{\theta})\,M\,R_z^{\otimes n}(\vec{\phi})\bigr)_{ij}.
  \label{eq:sjoj}
\end{equation}
In terms of these quantities, the fidelity is
\begin{equation}
  \bar F_d=\frac{\sum_j s_j+\bigl\lvert\sum_j o_j\bigr\rvert^2}{d(d+1)} .
  \label{eq:fbar_sjoj}
\end{equation}
With $p_j=\sum_i \lvert U_{ij}\rvert\,\bigl\lvert \bigl(R_z^{\otimes n}(\vec{\theta})MR_z^{\otimes n}(\vec{\phi})\bigr)_{ij}\bigr\rvert=\sum_i \lvert U_{ij}\rvert\,\lvert M_{ij}\rvert$,
the Cauchy--Schwarz and triangle inequalities order
\begin{equation}
  \sum_j\sqrt{s_j}\;\ge\;\sum_j p_j\;\ge\;\sum_j\lvert o_j\rvert\;\ge\;
  \Bigl\lvert\sum_j o_j\Bigr\rvert.
  \label{eq:fidchain}
\end{equation}
Squaring any term of Eq.~\eqref{eq:fidchain} and substituting it for
$\lvert\sum_j o_j\rvert^2$ in Eq.~\eqref{eq:fbar_sjoj} therefore gives an upper
bound on $\bar F_d$, evaluated from the same propagated states. The topmost bound, with $\sum_j\sqrt{s_j}$, equals unity when no population leaves the qubit subspace, and its difference from unity thus gives the \emph{leakage} error. The gaps between successive bounds, down to $\bar F_d$ itself, then attribute the remaining shortfall to three further sources: population reaching the wrong basis states (\emph{misrouting}), the relative phase within a single output superposition (\emph{internal phase error}), and the relative phase between different inputs (\emph{collective phase error}). The first three error channels are properties of individual initial states and can therefore be evaluated input by input, whereas the collective phase error compares the phases of the outputs of different initial states and has no per-input definition.

\subsection{Long-range gates as a test case}
\label{sec:longrange-test}
The gates of Sec.~\ref{sec:gate-engineering} are calibrated one pair at a time, with the spectators idle. In a circuit, however, gates act in sequence on a register in which any qubit may be excited or in a superposition, and it is not obvious that gates calibrated in isolation retain their fidelity when composed. We test this on a \sqiswap{03} gate between the two outer wells. Since the inter-well coupling decays with distance, the direct coupling between wells 0 and 3 is weak, so a direct gate requires a long hold time. We compare this direct long-range gate with routed versions that only use nearest-neighbor gates. In both cases we propagate all 16 qubit states through the complete schedule with all four wells correlated, i.e.\ with four orbitals per well.

\begin{figure}[htbp]
\centering
\includegraphics{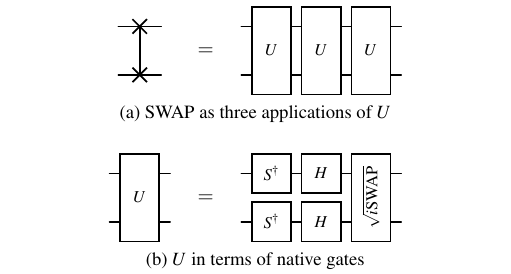}
\caption{Decomposition of a SWAP gate into the native $H$, $S^\dagger=e^{-i\pi/4}R_z(-\pi/2)$ and $\sqrt{i\mathrm{SWAP}}$ gates. The decomposition is valid up to a global phase $e^{-i\pi/4}$.}
\label{fig:swap_decomposition}
\end{figure}

For the routed gates, we assume access to two single-qubit gates, the Hadamard gate $H$ and the $R_z(\theta)$ gate with arbitrary rotation angle $\theta$. We neglect the execution time of these gates and model each as an ideal operation on its own well. Since the eigenstates at the idle point are not exactly products of single-well states, such a local operation introduces an error. We estimate that this error lowers the average gate fidelity of the routed gates by about 0.06--0.08, as shown in Sec.~\ref*{S-app:single-qubit} of the supplementary material. Together with the \sqiswap{} gates, these single-qubit gates allow SWAP gates to be constructed as in Fig.~\ref{fig:swap_decomposition}.

\begin{figure}[htbp]
\centering
\includegraphics{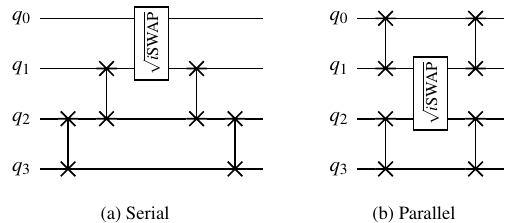}
\caption{Two different nearest-neighbor routings of the \sqiswap{03} gate: (a) serial, and (b) parallel. The synchronized variant uses the circuit in (b).}
\label{fig:circuits}
\end{figure}

Using the SWAP gates, we build two routings of \sqiswap{03}, shown in Fig.~\ref{fig:circuits}. In the \emph{serial} routing of Fig.~\ref{fig:circuits}(a), the state of well 3 is swapped via well 2 into well 1, entangled with well 0 by a \sqiswap{01} gate, and swapped back. In the \emph{parallel} routing of Fig.~\ref{fig:circuits}(b), the states of the outer wells are swapped simultaneously into the inner wells, entangled by a \sqiswap{12} gate, and swapped back. In both routings, each gate uses the control schedule and single-qubit $R_z$ rotation angles that maximize its own fidelity. The corresponding parameters are determined once, for the gate propagated on its own, and reused wherever the gate appears. For the parallel routing we additionally consider a \emph{synchronized} variant, in which the \sqiswap{01} and \sqiswap{23} gates use identical control schedules, so that the simultaneous SWAPs finish together. This common schedule is chosen by inspecting the fidelity landscapes of both gates for a point where each retains a high fidelity, at the cost of moving them away from their individual optima.

The direct gate is constructed with the same protocol as the nearest-neighbor gates, with a target operating point for the pair $(0,3)$ found as in Sec.~\ref{sec:op-points} and a control schedule optimized as in Sec.~\ref{sec:td-opt}.

\section{Results and discussion}
\label{sec:results}
We apply the multi-species MCTDHF method to the four-well register in two sets of simulations, which differ in the number of electrons per well. In the first set, presented in Sec.~\ref{sec:fourwell}, each well holds one electron. The intra-species interaction is then absent, and the method reduces to MCTDH. In this case, the calculations can be verified against an untruncated reference. We evaluate the gate optimization with mean-field spectators and the long-range gates of Sec.~\ref{sec:longrange-test}. In the second set, presented in Sec.~\ref{sec:threewell}, each well holds three spin-polarized electrons. Spin polarization accentuates the intra-well antisymmetrization, which makes full use of the MCTDHF framework, and we propagate the register with the multilayer ansatz of Sec.~\ref{sec:ml}. Here, we examine the nearest-neighbor gates and their convergence with the number of orbitals.

\subsection{Four-well register with one electron per well}
\label{sec:fourwell}
In the case of one electron per well, all 16 qubit states can be propagated on the joint $64^4$ grid, as described in App.~\ref{app:exact}. We use this untruncated propagation as a reference for MCTDH with four orbitals per well for the three nearest-neighbor gates, the direct \sqiswap{03} gate and the synchronized routing of Sec.~\ref{sec:longrange-test}. The register
fidelity $\bar F_{16}$ of MCTDH differs from the reference by at most
$1.7\times10^{-5}$ for the nearest-neighbor gates, by $8.5\times10^{-5}$ for
the direct \sqiswap{03} gate and by $4.7\times10^{-4}$ for the synchronized
routing, which is small compared with the infidelities of the gates
themselves. Four orbitals per well are therefore used for the full
propagations in the remainder of this section. In Sec.~\ref{sec:mf-res}, we
present the results of the mean-field two-qubit gate optimization, in which
the spectator wells are restricted to a single orbital. The results of the
long-range gates of Sec.~\ref{sec:longrange-test} are presented in
Sec.~\ref{sec:longrange-res}.

\subsubsection{Two-qubit gate optimization with mean-field spectators}
\label{sec:mf-res}
We sweep the ramp and hold times of the control schedule (Sec.~\ref{sec:td-opt}) over a grid derived from the static eigenstates of the operating points (Sec.~\ref*{S-app:gate-timing} of the supplementary material), for the three nearest-neighbor gates \sqiswap{01}, \sqiswap{12} and \sqiswap{23} and for the direct \sqiswap{03} gate. The full sweep results are shown in Fig.~\ref*{S-fig:1e_mf_sweep} of the supplementary material. In all four cases, the fidelity depends mostly on the hold time, which sets how far the exchange rotation proceeds. The ramp time matters much less and mainly trades gate duration against fidelity.

\begin{figure*}[ht!]
\centering
\includegraphics{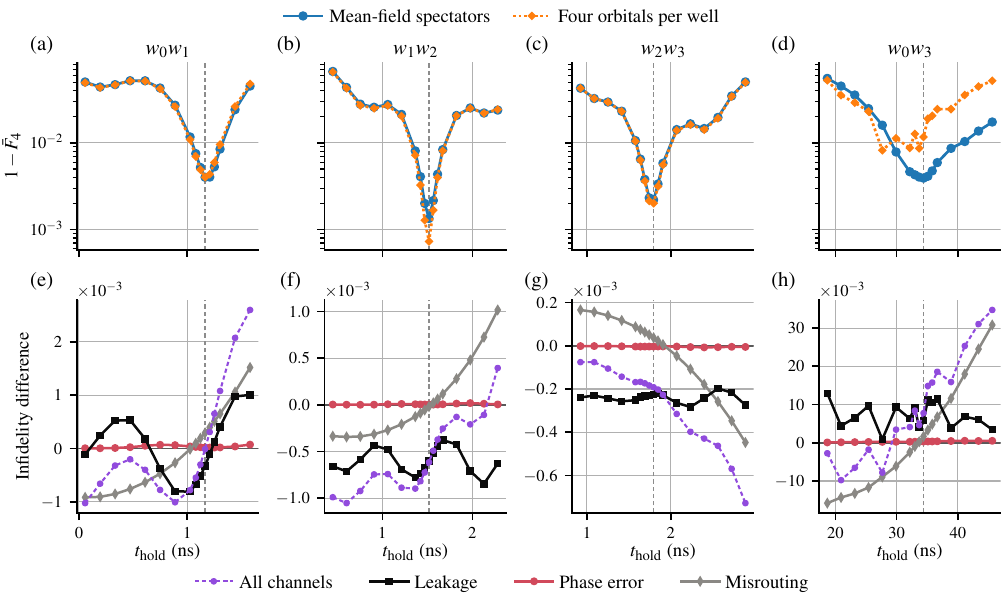}
\caption{Mean-field sweep against full four-orbital propagation for the three nearest-neighbor gates ($w_0w_1$, $w_1w_2$, $w_2w_3$) and for the direct \sqiswap{03} gate ($w_0w_3$). The mean-field sweep restricts the spectator wells to a single orbital, whereas the four-orbital propagation keeps four in every well. Panels (a)--(d) show the infidelity $1-\bar F_4$ against hold time at the optimal ramp time of each gate, $t_\mathrm{ramp}=\SI{1.31}{\nano\second}$, \SI{0.91}{\nano\second}, \SI{0.54}{\nano\second} and \SI{1.29}{\nano\second} for \sqiswap{01}, \sqiswap{12}, \sqiswap{23} and \sqiswap{03}, respectively. Panels (e)--(h) show the difference between the two sweeps, resolved by the error channels leakage, misrouting and phase error (Sec.~\ref{sec:fidelity}), as $\Delta X = X(\text{four orbitals})-X(\text{mean field})$. A positive $\Delta X$ means that the mean-field sweep under-reports channel $X$. The curve labeled \emph{All channels} is the sum of the three and equals $\Delta(1-\bar F_4)$. The optimum of the mean-field sweep is marked with a vertical dashed line in each panel. Note that the vertical scale of the lower panels differs between the gates.}
\label{fig:mf}
\end{figure*}

The mean-field hold-time sweeps at the best ramp times are compared with the full four-orbital propagation in Fig.~\ref{fig:mf}. At the optimum of each nearest-neighbor gate the four-orbital propagation gives $\bar F_4=0.99600$, $0.99927$ and $0.99799$ with gate durations of 3.78, 3.34 and \SI{2.88}{\nano\second} for \sqiswap{01}, \sqiswap{12} and \sqiswap{23}, respectively. Evaluated on the full register, i.e.\ for all four states of the spectators, the same gates give $\bar F_{16}=0.9413$, $0.9983$ and $0.9977$. The \sqiswap{01} gate thus depends strongly on the state of the spectators, which we return to in Sec.~\ref{sec:longrange-res}. The mean-field value of $\bar F_4$ is lower by between $1\times10^{-5}$ and $6\times10^{-4}$. The shape of the mean-field sweep agrees with the full propagation, and the minima of the infidelity $1-\bar F_4$ coincide, as shown in Fig.~\ref{fig:mf}(a)--(c). To identify the origin of the remaining difference, we decompose it into the error channels of Sec.~\ref{sec:fidelity}, as illustrated in Fig.~\ref{fig:mf}(e)--(g). The total difference is below $3\times10^{-3}$ at all hold times and arises from leakage and misrouting, whereas the phase error is the same in the mean-field and four-orbital propagations. The misrouting difference is small and changes sign close to the minimum, which means that the exchange rotation is completed at nearly the same hold time in both.

Figures~\ref{fig:mf}(d) and \ref{fig:mf}(h) show that the mean-field sweep is less reliable for the direct \sqiswap{03} gate. At the mean-field optimum, $t_{\mathrm{hold}}=\SI{34.4}{\nano\second}$, the mean-field sweep predicts $\bar F_4=0.99608$, but the four-orbital propagation gives $0.98832$. The mean-field value is therefore higher than the four-orbital one by $7.8\times10^{-3}$, whereas for the nearest-neighbor gates it is lower by at most $6\times10^{-4}$. Most of this difference is leakage. With four orbitals, the single-excitation inputs lose $0.4$--$0.5\,\%$ of their population, compared with $0.06\,\%$ in mean field. Moreover, the four-orbital leakage varies irregularly with the hold time. For the $|10\rangle$ input, the oscillation periods of the population loss match the one-body gaps between $w_3$ and the inner wells $w_1$ and $w_2$. These periods are shorter than the spacing of the sampled hold times, so that the oscillation is sampled at an essentially random phase. A possible cause is that the frequency of $w_3$ must cross those of the inner wells during the ramp in order to be resonant with $w_0$, which can excite the inner wells when they have more than one orbital. Individual four-orbital points are therefore uncertain by up to about $0.01$.

These results illustrate the tunability of the MCTDHF method. The orbital count can be chosen separately for each well, and for the nearest-neighbor gates a single orbital in the spectator wells is enough to locate the optimum at a reduced cost. However, the direct \sqiswap{03} gate shows that the truncation cannot be chosen independently of the physics, since a truncation that is accurate for one gate can fail for another if the dynamics involve excitations of the restricted wells. The orbital count therefore has to be motivated by the physical picture of the gate and verified against a calculation with more orbitals where possible.

\subsubsection{Direct and routed long-range gates}
\label{sec:longrange-res}
We propagate all 16 qubit states simultaneously through the circuits and control schedules of Sec.~\ref{sec:longrange-test}, using state averaging with four orbitals per well. The fidelity of the resulting matrix is evaluated against the \sqiswap{03} gate with the identity on the spectators. The direct gate reaches $\bar F_{16}=0.966$ with a gate duration of \SI{37.0}{\nano\second}. The serial routing gives $\bar F_{16}=0.873$ in \SI{41.1}{\nano\second}, the parallel routing $0.823$ in \SI{26.0}{\nano\second}, and the synchronized variant $0.856$ in \SI{23.1}{\nano\second}. Consequently, none of the routings improve the fidelity compared to the direct gate, but the parallel and synchronized routings reduce the gate duration by up to $38\,\%$.

\begin{figure*}[ht!]
    \centering
    \includegraphics{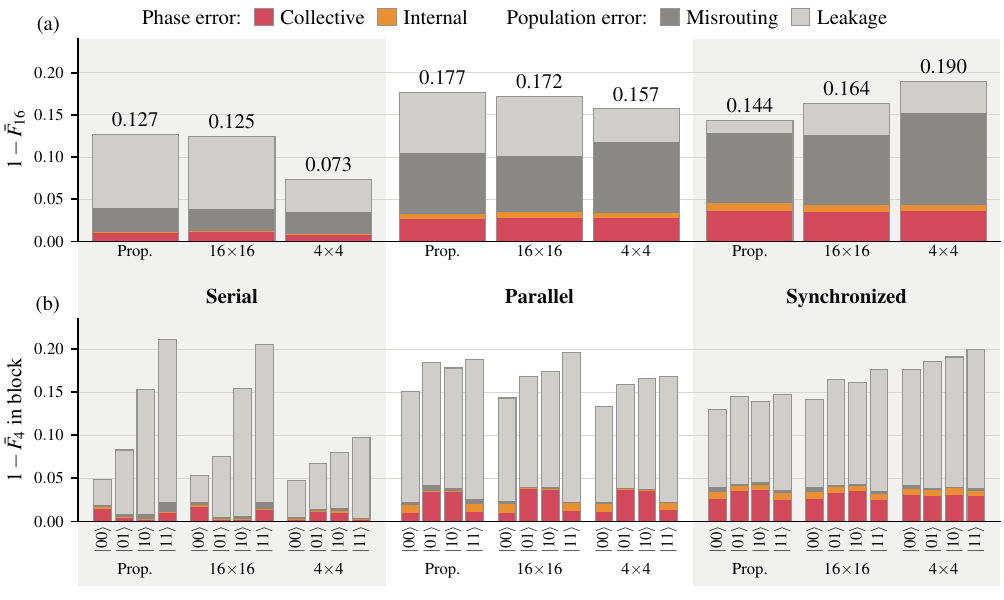}
    \caption{Infidelity decomposition of the routed \sqiswap{03} gate for three calculations of each circuit: propagation of the whole circuit (Prop.), and products of the individually propagated layers projected onto the full $16\times16$ qubit subspace ($16\times16$) or onto the $4\times4$ subspace of the active qubits with the spectators in their ground states ($4\times4$). (a) Register infidelity $1-\bar F_{16}$, decomposed into the error channels of Sec.~\ref{sec:fidelity}. The value above each bar is the total. (b) Infidelity $1-\bar F_4$ within each block of fixed initial spectator state $\ket{q_1q_2}$, evaluated with the $R_z$ angles of panel (a). Population that changes the spectator state leaves the block and counts as leakage in (b), but as misrouting in (a). Since the $R_z$ angles are not optimized per block, only the sum of the two phase errors is meaningful in (b).}
    \label{fig:methods_fid}
\end{figure*}

That said, the purpose of simulating the routed gates was primarily to test the MCTDHF method. We therefore compare the full propagation with two reduced models of the same circuits. In both, each layer of two-qubit gates in the circuits, either a single gate or two simultaneous gates, is propagated separately to obtain its gate matrix in the qubit subspace. The model denoted $4\times4$ projects the propagated states onto the $4\times4$ subspace of the active qubits with the spectators in their ground states. For two simultaneous gates, the two blocks are combined by a tensor product. The $16\times16$ model instead projects onto the full $16\times16$ qubit subspace, so that the dependence on the state of the spectators is retained. The circuits are then evaluated by matrix multiplication and the infidelity of the resulting routed \sqiswap{03} gates is decomposed into the four error channels of Sec.~\ref{sec:fidelity}. The decompositions are compared with that of the full propagation in Fig.~\ref{fig:methods_fid}.

Figure~\ref{fig:methods_fid}(a) shows that the reduced models act neither as an upper nor a lower bound on the infidelity of the circuit, since the propagation gives the highest infidelity for the serial and parallel routings, but the lowest infidelity for the synchronized routing. The difference in infidelity between the models mainly comes from how well the population errors are captured. The phase errors, in contrast, are similar in all three calculations, with the total phase error differing by less than $4\times10^{-3}$. This is because the phases are primarily determined by the coupling of the active pair during the gate, which is included in all three models. The largest difference is found for the serial routing, where the phases of the entangling \sqiswap{01} gate depend on the state of well~2, which the $4\times4$ model does not contain. The considerably larger differences in the population errors originate from two effects that are missing from one or both of the reduced models. First, interactions with the spectators can cause leakage out of the qubit subspace or misrouting into the wrong spectator configuration. The $4\times4$ model cannot capture this effect, but the $16\times16$ model can. Second, population that leaks out of the qubit subspace during one gate can partly return during a later gate, which neither model can describe. It should be noted that this return effect makes the fidelity of the circuits sensitive to how the $R_z$ phase corrections are applied. For each two-qubit gate on its own, a common phase can be distributed freely between the $R_z$ rotations before and after the gate without affecting its fidelity, as shown in Sec.~\ref*{S-app:rz-angles} of the supplementary material. However, this choice sets the phase with which returning population interferes with the population that remained in the qubit subspace, and thereby the fidelity of the circuit. All results presented here divide the common phase equally between the two layers of $R_z$ rotations.

In Fig.~\ref{fig:methods_fid}(b), the infidelity $1-\bar F_4$ for each initial state $\ket{q_1q_2}$ of the spectators, i.e.\ the qubits in wells 1 and 2, is decomposed into the same error channels as in panel (a). Since the $4\times4$ subspace used to calculate $\bar F_4$ is fixed by the initial state $\ket{q_1q_2}$, population transferred into any of the other three spectator states counts as leakage in panel (b), whereas it counts as misrouting in panel (a). For the serial routing, panel (b) shows that the leakage of the propagation is considerably larger in the blocks with $q_1=1$. In this routing, the state of $q_1$ is swapped into well 2 before the entangling \sqiswap{01} gate, so that well 2 is excited during the gate whenever $q_1=1$. This strongly increases the leakage of the \sqiswap{01} gate. Propagated on its own, the gate loses less than $0.5\,\%$ of the population of each input with well 2 in its ground state, but between $6$ and $22\,\%$ with well 2 excited. The $16\times16$ model captures this effect and reproduces the leakage of the propagation, whereas the $4\times4$ model, which only contains the gates with the spectators in their ground states, misses it and therefore underestimates the infidelity of the serial routing.

The dependence on the state of well 2 also explains why the \sqiswap{01} gate has the lowest $\bar F_{16}$ of the nearest-neighbor gates, as found in Sec.~\ref{sec:mf-res}. The serial routing applies it only once, whereas the parallel and synchronized routings apply it six times, since each of their two SWAPs between wells 0 and 1 consists of three \sqiswap{01} gates. In both of these routings, however, every \sqiswap{01} gate is applied simultaneously with a \sqiswap{23} gate. This increases the energy gap between wells 1 and 2 during the gate, which suppresses the leakage and makes it less dependent on the state of the spectators, as seen in Fig.~\ref{fig:methods_fid}(b). The parallel and synchronized routings are therefore only moderately less accurate than the serial routing, although they apply the least accurate gate six times instead of once.

Comparing the parallel and synchronized routings provides a direct test of this suppression, since they differ only in the control schedules of the simultaneous gates. In the parallel routing, the individually optimized \sqiswap{23} gate finishes \SI{0.9}{\nano\second} before the \sqiswap{01} gate, so that the leakage of the \sqiswap{01} gate is no longer suppressed during its last part. The synchronized routing decreases the leakage by keeping it suppressed during the whole gate, but instead increases the misrouting and phase errors since both gates are moved away from their individual optima. The $4\times4$ model lacks the leakage that the synchronization suppresses, but captures the larger misrouting and phase errors. It therefore predicts the synchronized routing to be less accurate than the parallel routing. Furthermore, the return effect discussed earlier is significant for the synchronized routing, but negligible for the parallel routing. Since neither of the reduced models captures the return effect, the propagation goes from giving the highest infidelity of the three calculations for the parallel routing to the lowest for the synchronized routing. In addition to improving the fidelity, the synchronized routing is \SI{2.9}{\nano\second} faster than the parallel routing.

\subsection{Four-well register with three electrons per well}
\label{sec:threewell}
\begin{figure}[ht]
    \centering
    \includegraphics{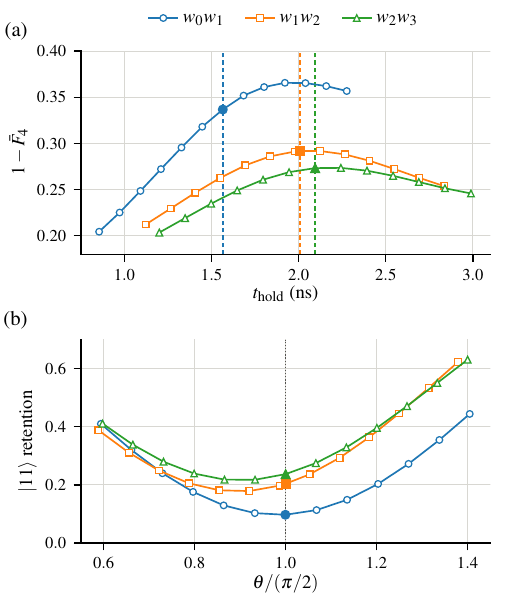}
    \caption{Hold-time sweeps of the \sqiswap{} gates with three electrons per well, using 22 orbitals per active well and mean-field spectators. The ramp times are $0.20$, $0.15$ and \SI{0.13}{\nano\second} for $w_0w_1$, $w_1w_2$ and $w_2w_3$, respectively. Filled markers, and the dashed lines in (a), mark the hold time that gives the optimal swap angle $\theta=\pi/2$. (a)~Infidelity $1-\bar F_4$ against the hold time. Unlike with one electron per well (Fig.~\ref{fig:mf}), the infidelity has no minimum at $\theta=\pi/2$. (b)~Retention of $\ket{11}$, i.e.\ its population after the gate, against the swap angle $\theta$. The optimal swap angle coincides closely with the lowest $\ket{11}$ retention for all three gates.}
    \label{fig:3e_qubit}
\end{figure}

In the case of three electrons per well, the \sqiswap{} gates of Sec.~\ref{sec:gate-engineering} fail. However, they are still a useful test case for the MCTDHF method, since the intra-well antisymmetrization is accentuated and the dynamics are more complex. We present a similar hold-time sweep as in Sec.~\ref{sec:mf-res} for the three nearest-neighbor gates in Fig.~\ref{fig:3e_qubit}(a). The sweep is performed with 22 orbitals per active well and mean-field spectators, at fixed ramp times of $0.20$, $0.15$ and \SI{0.13}{\nano\second} for \sqiswap{01}, \sqiswap{12} and \sqiswap{23}, respectively (see Sec.~\ref*{S-app:gate-timing} of the supplementary material for their choice). None of the gates reach an infidelity below $0.2$, and all of them have their lowest values at the shortest hold time. The reason for the poor fidelities is shown in Fig.~\ref{fig:3e_qubit}(b), where the population of the $\ket{11}$ state is plotted against the swap angle $\theta$, defined by the rotation of the single excitation, $\ket{10}\to\cos(\theta/2)\ket{10}-i\sin(\theta/2)\ket{01}$. The optimal swap angle, $\theta=\pi/2$, coincides closely with the lowest $\ket{11}$ retention for all three gates.

This is a consequence of the wells becoming nearly harmonic when there are three electrons in each well. The gate tunes the active wells into resonance, so that $\ket{10}$ and $\ket{01}$ become degenerate. In harmonic wells, this resonance also makes $\ket{11}$ degenerate with the states in which one well is doubly excited, since all three states hold two excitations. The operating point that drives the swap therefore also transfers population out of $\ket{11}$, and the lowest $\ket{11}$ retention lies close to the optimal swap angle. By contrast, the wells are strongly anharmonic with one electron per well. At the target operating point, the doubly excited states are detuned from $\ket{11}$ by about ten times the energy splitting $\Delta$ between the eigenstates $(\ket{10}\pm\ket{01})/\sqrt2$, so that $\ket{11}$ is retained. With three electrons per well, the qubit excitation is a center-of-mass oscillation of the electrons. The second excited energy level is an internal excitation, and the second quantum of the qubit excitation instead lies in the third excited energy level. The doubly excited states are therefore $\ket{30}$ and $\ket{03}$. While $\Delta$ is about the same as with one electron, the detuning of $\ket{30}$ and $\ket{03}$ from $\ket{11}$ drops to about $\Delta$ for all three gates. The detuning is smallest for the $w_0w_1$ gate, which also has the lowest $\ket{11}$ retention.

\begin{figure}[ht]
    \centering
    \includegraphics{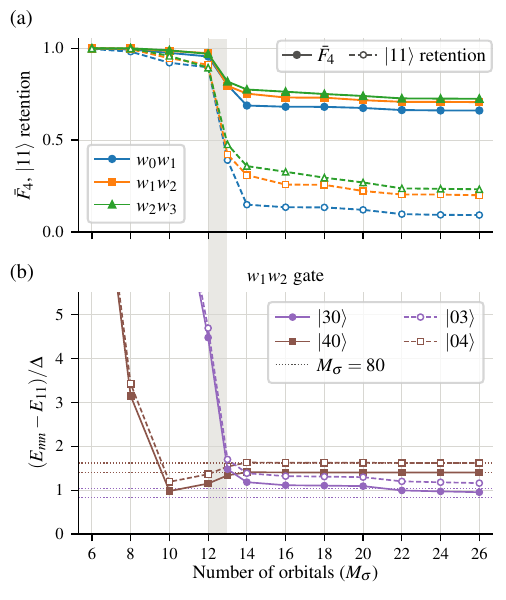}
    \caption{Convergence of the \sqiswap{} gates with three electrons per well and mean-field spectators, against the number of orbitals $M_\sigma$ per active well. Each gate keeps the ramp and hold times that give $\theta=\pi/2$ in Fig.~\ref{fig:3e_qubit}. The shaded band marks the step from 12 to 13 orbitals. (a)~$\bar F_4$ (solid) and retention of $\ket{11}$ (dashed) for the three gates. (b)~Energies of the states $\ket{30}$, $\ket{03}$, $\ket{40}$ and $\ket{04}$ of \sqiswap{12} relative to $\ket{11}$, in units of $\Delta$. The dotted lines show the same energies with all 80 orbitals of the grid. For each $M_\sigma$, the levels are labeled by their largest overlap with the 80-orbital levels, so that $\ket{30}$ and $\ket{03}$ are always the doubly excited states. The detuning of these states drops sharply between 12 and 13 orbitals, which explains the sudden collapse in $\ket{11}$ retention in panel (a).}
    \label{fig:3e_convergence}
\end{figure}

The detuning depends on how well the orbitals describe the doubly excited states, so we examine how the gates converge with the number of orbitals. We use the ramp and hold times that give $\theta=\pi/2$ in the 22-orbital sweeps of Fig.~\ref{fig:3e_qubit} and propagate the three nearest-neighbor gates with 6 to 26 orbitals per active well. The results are shown in Fig.~\ref{fig:3e_convergence}(a). Up to 12 orbitals, the gates appear to work, with $\bar F_4\geq0.955$ and at least $0.89$ of $\ket{11}$ retained. At 13 orbitals, the retention drops to $0.39$--$0.48$ and $\bar F_4$ to $0.80$--$0.82$ for all three gates, and both continue to fall up to 22 orbitals. Between 22 and 26 orbitals, $\bar F_4$ changes by at most $0.002$ per step and the retention by at most $0.004$. The swap angle, in contrast, is within $0.3\,\%$ of $\pi/2$ from 10 orbitals on. A calculation with too few orbitals therefore gets the swap right but misses the leakage, and reports a high-fidelity gate.

We investigate the abrupt drop in $\ket{11}$ retention at 13 orbitals by studying the energy spectrum of the active wells. For each $M_\sigma$, we take the time-evolved orbitals at the end of the ramp from the propagations of Fig.~\ref{fig:3e_convergence}(a). We diagonalize the Hamiltonian of each active well exactly within these orbitals. The other active well and the two spectator wells enter through the Coulomb potential of their ground-state densities. As a reference, we perform the same calculation on the full 80-point sine-DVR grid, i.e.\ $M_\sigma=80$. Since the energy order of the levels changes with $M_\sigma$, we label each level by its largest overlap with the 80-orbital levels. Figure~\ref{fig:3e_convergence}(b) shows the energies of $\ket{30}$ and $\ket{03}$ of \sqiswap{12} relative to $\ket{11}$, together with those of $\ket{40}$ and $\ket{04}$. Up to 12 orbitals, $\ket{30}$ and $\ket{03}$ lie more than $4\,\Delta$ above $\ket{11}$, and the retention in panel (a) stays high. At 13 orbitals, they drop to $1.5\,\Delta$ and $1.7\,\Delta$, at the same step at which the retention collapses. The states $\ket{40}$ and $\ket{04}$, by contrast, are already within $1.4\,\Delta$ of $\ket{11}$ at 10 and 12 orbitals without causing leakage. The fourth excited energy level is an internal excitation, so these states do not couple to $\ket{11}$. The abrupt drop in $\ket{11}$ retention is therefore caused by $\ket{30}$ and $\ket{03}$ coming close to resonance with $\ket{11}$ once the orbitals can describe them.

All calculations in this section treat the spectator wells at the mean-field level, i.e.\ with one orbital per electron, and use four SBS in each active well. We verify both choices at the hold times that give $\theta=\pi/2$. Treating the spectator wells with 16 orbitals and four SBS each changes the fidelities of the three gates by less than $10^{-2}$. Increasing the number of SBS per active well from four to six also changes them by less than $10^{-2}$. Mean-field spectators and four SBS per active well are therefore sufficient for the nearest-neighbor gates considered here.

These results illustrate the roles of the two truncation parameters of the multilayer ansatz. The fidelities depend only weakly on the number of SBS per active well, but strongly on the number of orbitals. With the 22 orbitals needed to converge the leakage of $\ket{11}$, each active well has $\binom{22}{3}=1540$ determinants. The single-layer ansatz of Eq.~\eqref{eq:ansatz} would then need $2\times10^{6}$ coefficients per state with mean-field spectators, and $7\times10^{11}$ with 16 orbitals in the spectator wells. In the multilayer ansatz, the top tensor instead has 16 and 256 elements per state, respectively. The number of SBS can be kept small since each well only populates a few of its many-body states. This is a general property of a quantum register, in which each qubit is meant to stay close to its two computational states. The multilayer ansatz is thus particularly well suited to systems of multiple qubits where the physics of the individual qubits might be hard to capture with a simple model.

\section{Conclusions and outlook}
\label{sec:conclusions}
In summary, we have shown that multi-species MCTDHF provides an accurate and compact description of gates in multi-qubit hardware, with \sqiswap{} gates in a model register of four anharmonic wells as a concrete example. In the case of one electron per well, four time-dependent orbitals in each well are sufficient to reproduce an untruncated propagation on the full 64-point grids, at an error well below the infidelities of the gates themselves. We have also shown that the nearest-neighbor gates can be optimized at a reduced cost with a single orbital in the spectator wells, whereas the same truncation overestimates the fidelity of the direct gate between the outermost qubits. The number of orbitals in each well therefore has to be motivated by the physics of the gate and verified against a calculation with more orbitals.

For the routed gates, the fidelity of a circuit cannot be accurately inferred from its gates propagated one at a time. The error of each gate depends on the state of the other qubits and on the population that earlier gates have moved out of the qubit subspace. A reliable prediction therefore requires propagating the full register through the complete circuit, which the compact MCTDH representation makes affordable. The propagation of a complete circuit also indicates how its errors can be reduced. Synchronizing the control schedules of the simultaneous \sqiswap{01} and \sqiswap{23} gates improves the fidelity compared with the parallel routing, since the leakage of the \sqiswap{01} gate then stays suppressed during the whole gate. Furthermore, the distribution of the single-qubit phase corrections between the layers of a circuit determines how returning population interferes. Both could be included in an optimization of the complete circuit, whereas the control schedules in this work were optimized gate by gate.

In the case of three electrons per well, the gates fail because the wells are nearly harmonic, so the operating point that drives the swap also brings $\ket{11}$ into resonance with the doubly excited states. This leakage is captured only with a sufficient number of orbitals, and a calculation with too few orbitals reports a high-fidelity gate. The fidelity does not decrease gradually with the number of orbitals, but collapses at a single step, where the orbitals first describe the doubly excited states well enough to bring them close to resonance. With fewer orbitals, the doubly excited states lie too far from $\ket{11}$ to be populated, and the gate appears to work. This can be detected without a larger propagation, by diagonalizing each active well in the propagated orbitals and in the full basis. Capturing the leakage requires many orbitals, which is affordable because each well populates only a few of its many-body states. Hence, the multilayer ansatz needs only a few species basis states per well. This is a general property of a quantum register, which makes the multilayer ansatz well suited to qubits whose internal structure is not captured by a simple model.

The calculations presented here are limited to a closed system, with the qubits encoded in the motional states of the electrons. The framework, however, only assumes that the qubits are distinguishable subsystems of interacting particles. It can therefore be extended to qubits encoded in the spin. In spin--charge-coupled qubits, such as those proposed for electrons on helium~\cite{jenningsQuantumComputingUsing2024,dykmanSpinDynamicsQuantum2023}, the spin of each electron is coupled to its motion, and the wells remain distinguishable species. The extension then amounts to spin orbitals and a spin-dependent one-body operator. In singlet--triplet qubits~\cite{burkardSemiconductorSpinQubits2023}, each qubit consists of two electrons in a tunnel-coupled double quantum dot, so that the distinguishable subsystems are the double dots rather than the individual wells. Furthermore, MCTDH has been generalized to the propagation of density operators~\cite{raabMulticonfigurationTimedependentHartree1999} and formulated for master equations of Lindblad form, for any mixture of species~\cite{kvaalMulticonfigurationalTimedependentHartree2011}. Combining this formulation with the present framework would allow decoherence and relaxation to be treated in the same calculation as the leakage, misrouting and phase errors. The errors caused by the control could then be separated from those caused by the environment.

\section*{Supplementary Material}
See the supplementary material for derivations of the anharmonicity of the confining potential and the effective quasi-one-dimensional interaction, additional details on the construction and optimization of the operating points, the single-qubit gates of the routed circuits, and the timing windows, sweeps and final control schedules of the two-qubit gates.

\begin{acknowledgments}
The author thanks Professor Morten Hjorth-Jensen for guidance throughout this work.
This project has received funding from the European Union's Horizon 2020 research and innovation program under the Marie Sk\l{}odowska-Curie Grant Agreement No.~945371.
\end{acknowledgments}

\section*{Author Declarations}
\subsection*{Conflict of Interest}
The author has no conflicts to disclose.

\section*{Data Availability}
The data that support the findings of this study are openly available in
Zenodo at \url{https://doi.org/10.5281/zenodo.24593453}.

\appendix
\section{Direct evaluation of the coefficient equation of motion}
\label{app:eom}

Because the orbitals move, the Hamiltonian matrix in Eq.~\eqref{eq:ci}
changes at every step and is never formed. We evaluate its action
directly~\cite{hochstuhlTimedependentMulticonfigurationMethods2014}, sorting
$(HC)_I = \sum_J \langle I|\hat H(t)|J\rangle C_J$ by which species the
excitation operators act on, $(HC)_I = \sum_\sigma (HC)^{\sigma}_I +
\sum_{\sigma<\tau} (HC)^{\sigma\tau}_I$, with
\begin{align}
  (HC)^{\sigma}_{I}
  &= \sum_{J_\sigma} H^{\sigma}_{I_\sigma J_\sigma}\,
     C_{\ldots J_\sigma\ldots},
  \label{eq:sigma1}\\
  (HC)^{\sigma\tau}_{I}
  &= \sum_{pqrs} g^{\sigma\tau}_{pqrs}
     \sum_{J_\sigma J_\tau}
     \langle I_\sigma|\hat a^{\dagger}_{p\sigma}
       \hat a_{q\sigma}|J_\sigma\rangle
  \nonumber\\
  &\qquad \times
     \langle I_\tau|\hat a^{\dagger}_{r\tau}
       \hat a_{s\tau}|J_\tau\rangle\,
     C_{\ldots J_\sigma\ldots J_\tau\ldots},
  \label{eq:sigma2}
\end{align}
where $C_{\ldots J_\sigma\ldots}$ denotes the coefficient with the index $I_\sigma$
replaced by $J_\sigma$, and $C_{\ldots J_\sigma\ldots J_\tau\ldots}$ the one with
$I_\sigma$ and $I_\tau$ replaced by $J_\sigma$ and $J_\tau$, all other indices being
those of $I$. $H^{\sigma}$ is the within-species Hamiltonian block, containing
$h^{\sigma}_{pq}$ and $g^{\sigma\sigma}_{pqrs}$. No term of Eq.~\eqref{eq:ham}
touches more than two of the $K$ configuration indices, so each contribution
is a contraction of the coefficient tensor along one or two of its modes.
Evaluating Eq.~\eqref{eq:sigma2} for every pair of full configurations at
once would cost as much as multiplying by a dense Hamiltonian matrix,
quadratic in the size of the coefficient tensor. Instead we sum over
$J_\tau$ first, at fixed $J_\sigma$, and only then sum over
$J_\sigma$; since each step varies only one species' index, the combined
cost is linear in the size of the coefficient tensor.

\section{Numerical details of the propagation}
\label{app:propagation}
\subsection{One electron per well}
The real-time MCTDH propagations of Sec.~\ref{sec:fourwell} use a midpoint variant of the second-order constant-mean-field (CMF) scheme~\cite{beckEfficientRobustIntegration1997,beckMulticonfigurationTimedependentHartree2000} with a fixed step $\Delta t$. We use $\Delta t=\SI{0.80}{\pico\second}$ for the nearest-neighbor gates propagated on their own, and \SI{0.32}{\pico\second} for the direct \sqiswap{03} gate and the routed circuits. This choice is motivated in Sec.~\ref*{S-app:propagation-tests} of the supplementary material. The scheme propagates the configuration coefficients $C_I$ and the orbitals separately. The coefficients are propagated with the orbitals held fixed. The orbitals are propagated with the density matrices, the mean-field operators $\hat g^{\sigma\tau}_{rs}$ and the projector $\hat P^{\sigma}$ held fixed. Both Eq.~\eqref{eq:ci} and Eq.~\eqref{eq:bdot} are then linear. In each step, the coefficients are first propagated over a half step in the orbitals at the beginning of the step. The orbitals are then propagated over the full step, with the fixed quantities evaluated at the midpoint from orbitals predicted by a preliminary half step. Finally, the coefficients are propagated over the second half step in the orbitals at the end of the step. The inverse density matrix is regularized as $(D^\sigma+\epsilon\,\openone)^{-1}$ with $\epsilon=10^{-6}$.

With four orbitals per well, each state has $4^4=256$ coefficients $C_I$, and the four orbitals of each well are expanded in 64 sine-DVR functions, giving 256 parameters $b^\sigma_{kn}$. The Hamiltonian matrix of Eq.~\eqref{eq:ci} and the matrix on the right-hand side of Eq.~\eqref{eq:bdot} are therefore small enough to be formed explicitly, so we propagate both equations exactly by their matrix exponentials. The only remaining error of the time integration is the CMF error, which arises because the quantities held fixed in the CMF scheme change continuously in the exact dynamics. For the second-order scheme this error is $\mathcal{O}(\Delta t^3)$ per step.

Since the matrix on the right-hand side of Eq.~\eqref{eq:bdot} is not Hermitian, its exponential preserves the orthonormality of the orbitals only to $\mathcal{O}(\Delta t^2)$ per step, and the error accumulates over the propagation. Every \SI{1.59}{\pico\second}, we therefore re-orthonormalize the orbitals of each well by a QR decomposition $b^\sigma=Q^\sigma R^\sigma$, replacing $b^\sigma$ by $Q^\sigma$ and absorbing $R^\sigma$ into the coefficients (Sec.~\ref{sec:eom}), which leaves $|\Psi\rangle$ unchanged. In a test propagation of \SI{17}{\nano\second} with $\Delta t=\SI{0.16}{\pico\second}$, the norm of $|\Psi\rangle$ changes by an amount of order one without the re-orthonormalization, and by only $5\times10^{-8}$ with it. For the nearest-neighbor gates of Sec.~\ref{sec:mf-res}, propagated with $\Delta t=\SI{0.80}{\pico\second}$, it changes by at most $3\times10^{-5}$.

\subsection{Three electrons per well}
\label{app:propagation-3e}
The multilayer propagations of Sec.~\ref{sec:threewell} use the corresponding second-order CMF scheme for the three layers of the multilayer ansatz, with a fixed step $\Delta t=\SI{0.16}{\pico\second}$. The scheme propagates the top tensor $A^N$, the SBS coefficients $B^\sigma$ and the orbitals separately, each with the other two held fixed. In each step, the top tensor is first propagated over a half step in the SBS and orbitals at the beginning of the step. The SBS coefficients are then propagated over a half step, with $\rho^\sigma$ and the single-hole functions of Eq.~\eqref{eq:ml-rho} evaluated from the top tensor at the midpoint. Next, the orbitals are propagated over the full step, with the density matrices evaluated from the top tensor and the SBS coefficients at the midpoint. Finally, the SBS coefficients and the top tensor are propagated over a second half step in the orbitals at the end of the step.

The top tensor has at most $4^4=256$ elements per state. As with one electron per well, the Hamiltonian matrix of Eq.~\eqref{eq:ml-A} is therefore small enough to be formed explicitly, so we propagate the top tensor exactly by its matrix exponential. Equations~\eqref{eq:ml-B} and~\eqref{eq:bdot} are too large for this, and we propagate them with the Crank--Nicolson scheme, solving the linear system iteratively with BiCGSTAB~\cite{vandervorstBiCGSTABFastSmoothly1992}. To reduce the cost, the linear systems are solved in mixed precision. Each system is solved for the increment of the coefficients over the sub-step, and not for the coefficients themselves. The right-hand side is evaluated in double precision, and the system is then solved in single precision to a relative tolerance of $10^{-4}$. Before the increment is added to the coefficients, the projector is applied to it once more in double precision. Without this, the single-precision error of the projection is amplified by the regularized inverse density matrices, and the orthonormality is lost within a few steps. Using mixed precision halves the cost of a step.

Within a step, each half step of the SBS coefficients is divided into 16 sub-steps of \SI{5.0}{\femto\second}, and the orbital step into 4 sub-steps of \SI{40}{\femto\second}, with the projector updated at the start of each sub-step. The error of the time integration is set by the lengths of these sub-steps and not by $\Delta t$. As with one electron per well, we re-orthonormalize the orbitals and the SBS by QR decompositions every \SI{1.59}{\pico\second}, which leaves $|\Psi_N\rangle$ unchanged. In addition, both are orthonormalized after every step by Löwdin symmetric orthogonalization~\cite{lowdinNonOrthogonalityProblemConnected1950}, which does not rotate the orbitals or the SBS among themselves. Unlike the QR re-orthonormalization, this changes the state, since it restores the norm of $|\Psi_N\rangle$, which the sub-steps do not conserve exactly. The norm is then conserved by construction and is not a measure of the integration error.
The integrator settings are verified in Sec.~\ref*{S-app:propagation-tests} of the supplementary material.

\subsection{Primitive grid convergence}
\label{app:grid}
We verify that the calculations are converged with respect to the number of sine-DVR grid points. With one electron per well, we test the grid on a pair of neighboring wells. The four lowest eigenvalues with 48 and with 64 points per well agree with those with 96 points to more than twelve digits, so the 64-point grid is converged. With three electrons per well, the grid also has to resolve the effective interaction $V_{\mathrm{eff}}$ at short separations. The three-electron results are most sensitive to the detunings of the leakage states $\ket{30}$ and $\ket{03}$ from $\ket{11}$ (Sec.~\ref{sec:threewell}), so we test the grid on these. We calculate the energy levels of each active well in isolation at the target operating points, and obtain the detunings from them. The calculation is performed using full CI on sine-DVR grids with 64, 80, 96 and 128 points per well. The 128-point values are taken as the reference. With 64 points, the detunings differ from the reference by up to $0.48\,\Delta$, and with 80 points by at most $0.003\,\Delta$. We therefore use 80 grid points for the calculations with three electrons per well. For the \sqiswap{12} gate, this raises the $\ket{11}$ retention by $0.15$--$0.18$ compared with 64 points. Since $\Delta$ changes by at most $0.06\,\%$ between 64 and 80 points, we keep the operating points optimized with 64 points and re-evaluate only the cost functions and populations with 80 points. The resulting values are listed in Tables~\ref*{S-tab:idle_point_3e} and~\ref*{S-tab:target_points_3e} of the supplementary material.

\section{Untruncated reference propagation}
\label{app:exact}
\begin{table*}
\caption{Four-orbital MCTDH against the untruncated reference using all 16
inputs. Each gate is scored against its own $\sqrt{i\mathrm{SWAP}}$ target and
the synchronized routing (Sync.) against \sqiswap{03}. Each entry is
$\Delta X=X(\text{reference})-X(\text{MCTDH})$ for the infidelity
$1-\bar F_{16}$ and its error channels (Sec.~\ref{sec:fidelity}), with the
reference extrapolated to zero step size. The phase error is the sum of the internal and collective phase errors. The three channels add up to
$\Delta(1-\bar F_{16})$. The last row is the residual step error of the
reference in $\bar F_{16}$ after extrapolation. The residual errors of the
individual channels are of similar size, except for the synchronized
routing, where they reach $1.0\times10^{-4}$ in the misrouting and
$8\times10^{-5}$ in the phase error. The difference in the phase error of the
synchronized routing is therefore not resolved.}
\label{tab:exact}
\begin{ruledtabular}
\begin{tabular}{lccccc}
 & \sqiswap{01} & \sqiswap{12} & \sqiswap{23} & \sqiswap{03} & Sync. \\[3pt]
\hline\noalign{\vskip3pt}
$t_\mathrm{gate}$ (ns) & 3.8 & 3.3 & 2.9 & 37.0 & 23.1 \\
$\Delta(1-\bar F_{16})$ & $1.7\!\times\!10^{-5}$ & $1.9\!\times\!10^{-6}$ & $2.4\!\times\!10^{-7}$ & $-8.5\!\times\!10^{-5}$ & $-4.7\!\times\!10^{-4}$ \\
Leakage & $1.7\!\times\!10^{-5}$ & $1.8\!\times\!10^{-6}$ & $2.0\!\times\!10^{-7}$ & $-7.2\!\times\!10^{-5}$ & $-2.6\!\times\!10^{-4}$ \\
Misrouting & $<10^{-7}$ & $<10^{-7}$ & $<10^{-7}$ & $-1.1\!\times\!10^{-6}$ & $-1.7\!\times\!10^{-4}$ \\
Phase error & $<10^{-7}$ & $<10^{-7}$ & $<10^{-7}$ & $-1.2\!\times\!10^{-5}$ & $-4.0\!\times\!10^{-5}$ \\[3pt]
\hline\noalign{\vskip3pt}
Ref.\ error & $4.3\!\times\!10^{-7}$ & $8.2\!\times\!10^{-8}$ & $1.8\!\times\!10^{-7}$ & $2.6\!\times\!10^{-6}$ & $2.9\!\times\!10^{-5}$ \\
\end{tabular}
\end{ruledtabular}
\end{table*}
For one electron per well, the wavefunction on the joint $64^4$ sine-DVR grid
has $1.7\times10^{7}$ amplitudes, so the 16 qubit states can be
propagated without orbital truncation, which provides a reference for the MCTDH
calculations. The reference uses the same grid, control schedules, qubit
orbitals and $R_z$ rotations as MCTDH, so that the two differ only in the
orbital truncation and the time integration. It is propagated in double
precision with a fourth-order splitting scheme at two step sizes:
\SI{3.98}{\pico\second} and \SI{1.99}{\pico\second} for the nearest-neighbor
gates, and \SI{7.96}{\pico\second} and \SI{3.98}{\pico\second} for the direct
\sqiswap{03} gate and the synchronized routing. The two results are
extrapolated to zero step size by Richardson extrapolation for an
$\mathcal{O}(\Delta t^4)$ error, and the difference between the extrapolated
value and the value at the smaller step is taken as the residual error of the
reference. To check that the step error is dominated by its leading
$\mathcal{O}(\Delta t^4)$ term, the nearest-neighbor gates are also propagated
at \SI{7.96}{\pico\second}. Extrapolating from the two larger steps, as for the \sqiswap{03} gate and the
routing, then gives the same $\bar F_{16}$ as extrapolating from the two
smaller ones, to within $3\times10^{-6}$.

The comparison is summarized in Table~\ref{tab:exact}. MCTDH with four
orbitals per well reproduces the reference closely in all five calculations.
The infidelity differs by $1.7\times10^{-5}$ or less for the nearest-neighbor
gates, $8.5\times10^{-5}$ for the direct \sqiswap{03} gate and
$4.7\times10^{-4}$ for the synchronized circuit, so the difference grows with
the duration and complexity of the propagation. Even for the circuit, it is
only $0.3\,\%$ of the infidelity of the reference. For the \sqiswap{23} gate,
the difference is comparable to the residual error of the reference and is
therefore not resolved. The error channels agree equally well. The phase error
is reproduced to within $10^{-7}$ for the nearest-neighbor gates and
$1.2\times10^{-5}$ for the direct gate, and to within the residual error of the
reference for the circuit. The remaining differences lie mainly in the leakage
and, for the circuit, also in the misrouting. Even there, they amount to
$1.8\,\%$ of the leakage and $0.2\,\%$ of the misrouting of the reference. As
for the mean-field spectators in Sec.~\ref{sec:mf-res}, the truncation thus
affects the populations more than the phases. The differences also include the time-step error of the CMF integration in the
MCTDH calculations (App.~\ref{app:propagation}). For
comparison, the product of the individually propagated gates, projected onto
the full 16-dimensional qubit subspace (the $16\times16$ model of
Sec.~\ref{sec:longrange-res}), differs from the reference by
$2.1\times10^{-2}$ in $\bar F_{16}$ for the same circuit, 45 times more than
MCTDH.

\renewcommand{\refname}{References}
\bibliography{references}

\end{document}


\preprint{AIP/123-QED}

\title[Supplementary Material: Multi-species MCTDHF for gate simulations in multi-qubit hardware]{Supplementary Material: Multi-species MCTDHF for gate simulations in multi-qubit hardware}
\author{Oskar Leinonen}
\email{oskar.leinonen@fys.uio.no}
\affiliation{ 
Department of Physics, University of Oslo, Oslo N-0316, Norway
}

\date{\today}

\maketitle
This supplementary material collects details of the model system and of the gate protocol that are not needed to follow the main text. Sections~\ref{app:anharm} and~\ref{app:veff} concern the model: the anharmonicity parameter of the confining potential, and the effective one-dimensional interaction between electrons in the same well. Section~\ref{app:op-points} describes how the idle and target operating points are constructed and optimized, and lists the resulting parameters. Section~\ref{app:single-qubit} describes the single-qubit gates of the routed circuits and the error they introduce. Section~\ref{app:gate-timing} derives the ramp and hold times over which the control schedules are swept, and presents the results of the sweeps. Section, equation, figure and table numbers without the prefix S refer to the main text.

\section{Anharmonicity of the confining potential}
\label{app:anharm}
Consider a single well, dropping the species and center labels for
clarity,
\begin{equation}
  \hat H = \frac{\hat p^2}{2} + \frac{1}{2}\omega^2\hat x^2
  + \frac{\beta\omega^2}{3}\hat x^4,
\end{equation}
and treat the quartic term as a perturbation to the harmonic oscillator
$\hat H_0=\hat p^2/2+\omega^2\hat x^2/2$, with eigenstates $|n\rangle$ and
$\hat x=(a+a^{\dagger})/\sqrt{2\omega}$. The quartic matrix element is the
standard result~\cite{LANDAU1977133}
\begin{equation}
  \langle n|\hat x^4|n\rangle = \frac{3}{4\omega^2}(2n^2+2n+1),
\end{equation}
so that first-order perturbation theory gives
\begin{equation}
  E_n^{(1)} = \frac{\beta\omega^2}{3}\langle n|\hat x^4|n\rangle
  = \frac{\beta}{4}(2n^2+2n+1).
\end{equation}
The energy levels are therefore 
\begin{equation}
    E_n = \omega(n+\tfrac12) +
\frac{\beta}{4}(2n^2+2n+1),
\end{equation}
to first order. Defining the anharmonicity as
the difference between successive level spacings,
$(E_2-E_1)-(E_1-E_0)$, gives exactly $\beta$: $(E_1-E_0)=\omega+\beta$ and
$(E_2-E_1)=\omega+2\beta$, so the harmonic spacing $\omega$ cancels and the difference is $\beta$. In other words, defining the quartic coefficient as $\beta\omega^2/3$ makes $\beta$ the first-order anharmonicity of each well, independent of $\omega$.

\section{Effective quasi-one-dimensional interaction}
\label{app:veff}

The Hamiltonian of Eq.~\eqref{eq:ham-fourwell} is one-dimensional. For
electrons in different wells the bare Coulomb potential $\kappa/|x-x'|$ is
unproblematic, since the grids are disjoint. For two electrons in the same
well, however, the Coulomb potential diverges at zero separation, and its matrix elements
in the primitive basis are ill-defined. We therefore replace it by an
effective interaction $V_{\mathrm{eff}}$, obtained by giving each electron
two auxiliary transverse coordinates $y$ and $z$ and averaging the
three-dimensional Coulomb potential over a fixed transverse state~\cite{bednarekEffectiveInteractionCharge2003}. Writing the wavefunction of an electron pair as
\begin{equation}
  \Psi(\mathbf r_1,\mathbf r_2)=\psi(x_1,x_2)\,
  \eta(y_1)\eta(y_2)\,\zeta(z_1)\zeta(z_2),
  \label{eq:frozen}
\end{equation}
and integrating out the transverse coordinates, we obtain
\begin{equation}
  V_{\mathrm{eff}}(x)=\kappa\int 
  \frac{|\eta(y_1)\eta(y_2)\,\zeta(z_1)\zeta(z_2)|^2}
       {\sqrt{x^2+(y_1-y_2)^2+(z_1-z_2)^2}} \dd^2y\,\dd^2z\,,
  \label{eq:veff_proj}
\end{equation}
where $x=x_1-x_2$ is the axial separation and
$\dd^2y=\dd y_1\,\dd y_2$, $\dd^2z=\dd z_1\,\dd z_2$. For the transverse states we take
$\eta$ as the ground state of a harmonic oscillator with frequency $\omega_y/2\pi=\SI{30}{\giga\hertz}$, and $\zeta$ as the ground state of an electron above a
liquid-helium surface. There the electron is bound by its image potential
and kept out of the liquid by a hard barrier at $z=0$, giving
$\zeta(z)\propto z\,e^{-z/r_B}$ ($z\ge0$) with effective Bohr radius
$r_B=4a_0(\varepsilon_{\mathrm{He}}+1)/(\varepsilon_{\mathrm{He}}-1)
\approx\SI{7.6}{\nano\meter}$, where $a_0$ is the Bohr radius and
$\varepsilon_{\mathrm{He}}$ is the dielectric constant of liquid helium~\cite{jenningsQuantumComputingUsing2024}. Both electrons occupy the same transverse orbital $\eta(y)\zeta(z)$, so the
product $\eta(y_1)\eta(y_2)\zeta(z_1)\zeta(z_2)$ is unchanged when the
electrons are exchanged. The antisymmetry within a well is therefore carried
entirely by $\psi$, as in the one-dimensional model.

The integrand of Eq.~\eqref{eq:veff_proj} depends on the transverse
coordinates only through the relative displacements $u=y_1-y_2$ and
$w=z_1-z_2$. Substituting $y_1=y_2+u$ and $z_1=z_2+w$ and integrating over
$y_2$ and $z_2$ gives
\begin{equation}
  V_{\mathrm{eff}}(x)=\kappa\int
  \frac{P_y(u)\,P_z(w)}{\sqrt{x^2+u^2+w^2}}\,\dd u\,\dd w,
  \label{eq:veff_avg}
\end{equation}
where
\begin{equation}
P_y(u)=\int |\eta(t+u)|^2\,|\eta(t)|^2\,\dd t,\qquad
P_z(w)=\int |\zeta(t+w)|^2\,|\zeta(t)|^2\,\dd t
\label{eq:P_def}
\end{equation}
are the distributions of the relative transverse coordinates. Each is the
autocorrelation of the corresponding single-electron density, and is
therefore even in its argument and normalized.

We now evaluate the two distributions, starting with the surface-normal direction. The density of $\zeta$ is $|\zeta(z)|^2=\tfrac12 a^3z^2e^{-az}$ for $z\ge0$, with $a=2/r_B$. Since
the density vanishes for $z<0$, the integral in Eq.~\eqref{eq:P_def} runs
over $t\ge0$, and for $w\ge0$
\begin{equation}
P_z(w)=\frac{a^6}{4}\,e^{-aw}\int_0^\infty t^2(t+w)^2e^{-2at}\,\dd t
=\frac{a}{16}\bigl(3+3aw+a^2w^2\bigr)e^{-aw},
\label{eq:Pz}
\end{equation}
where we used $\int_0^\infty t^ne^{-2at}\,\dd t=n!/(2a)^{n+1}$. For $w<0$,
$P_z(w)=P_z(|w|)$.

Turning to the harmonic direction, $|\eta(y)|^2$ is a Gaussian of variance
$1/(2\omega_y)$. The autocorrelation of a Gaussian is again a
Gaussian, with twice the variance, so
\begin{equation}
  P_y(u)=\frac{e^{-u^2/2\sigma_u^2}}{\sqrt{2\pi}\,\sigma_u},
  \qquad \sigma_u^2=\frac{1}{\omega_y}.
  \label{eq:Py}
\end{equation}
Because $P_y$ is Gaussian, the $u$-integral in Eq.~\eqref{eq:veff_avg} can be done analytically at fixed $w$. With $c^2=x^2+w^2$ and the substitution $u=c\sinh\theta$,
\begin{equation}
\int\frac{P_y(u)}{\sqrt{u^2+c^2}}\,\dd u
=\frac{1}{\sqrt{2\pi}\,\sigma_u}\int_{-\infty}^{\infty}
e^{-2q\sinh^2\theta}\,\dd\theta
=\frac{e^{q}}{\sqrt{2\pi}\,\sigma_u}\int_{0}^{\infty}
e^{-q\cosh\phi}\,\dd\phi
=\frac{e^{q}K_0(q)}{\sqrt{2\pi}\,\sigma_u},
\label{eq:veff_bessel}
\end{equation}
where $q=c^2/4\sigma_u^2$. In the second step we used
$2\sinh^2\theta=\cosh2\theta-1$ and set $\phi=2\theta$, and the remaining
integral is the standard integral representation of the modified Bessel
function $K_0$.

Inserting Eqs.~\eqref{eq:Pz} and \eqref{eq:veff_bessel} into
Eq.~\eqref{eq:veff_avg}, using that the integrand is even in $w$, and
changing variables to $s=aw$, we obtain
\begin{equation}
  V_{\mathrm{eff}}(x)=\frac{\kappa}{\sqrt{2\pi}\,\sigma_u}
  \int_0^\infty \frac{3+3s+s^2}{8}\,e^{-s}\,e^{q}K_0(q)\,\dd s,
  \label{eq:veff_final}
\end{equation}
with $q=(x^2+s^2/a^2)/4\sigma_u^2$. We evaluate this one-dimensional
integral by numerical quadrature on the grid of axial separations. For
$|x|\gg\sigma_u,r_B$, the asymptotic form $e^qK_0(q)\to\sqrt{\pi/2q}$
recovers the bare interaction $\kappa/|x|$. At $x=0$, $K_0$ has only a
logarithmic singularity at $s=0$, which is integrable, so
$V_{\mathrm{eff}}(0)$ is finite.

\section{Constructing and optimizing the operating points}
\label{app:op-points}
In this section we present the details of the operating-point optimization. The procedure is different between the case of one electron per well and three electrons per well, as described below.
\subsection{One electron per well}
The operating-point optimization can be divided into two parts. The first is to calculate the eigenstates of the system given a set of well parameters $\boldsymbol{\lambda}$, and the second is to use these eigenstates to locate the operating points that satisfy the desired conditions for the idle and target operating points. 

\subsubsection{Constructing the eigenstates}
\label{app:1e-construct}
In the case of one electron per well, the eigenstates of the system are obtained from imaginary-time evolution (ITE) of the MCTDH(F) equations of motion with four orbitals per well. We initialize 50 states by diagonalizing the uncoupled single-well Hamiltonians and constructing product states from the resulting single-well eigenstates. Because each such state occupies a single configuration exactly, we perturb the configuration coefficients with a small random admixture of amplitude $10^{-3}$ and re-orthonormalize before propagation, to facilitate the ITE convergence into the eigenstates of the coupled system. These initial states are then propagated simultaneously using state-averaged ITE (Sec.~\ref{sec:stateavg}), with weights $w_N=1/3$ for the ground state, $w_N=1/6$ for the next four states, and $w_N=0$ for all higher states (Eq.~\eqref{eq:densities-sa}). 

At each ITE step, the configuration coefficients of the propagated states are diagonalized within the instantaneous MCTDHF basis, as in the improved relaxation method~\cite{meyerCalculationSelectivePopulation2006}, so the states remain distinct and energy-ordered throughout the propagation.

The equations of motion have a residual gauge freedom: a joint rotation of the orbitals and the configuration coefficients leaves the physical state unchanged. We fix the gauge by diagonalizing the state-averaged one-particle density matrix, and then work in its eigenstates, which form the natural-orbital basis.

\subsubsection{Locating operating points}
\label{app:1e-locate}
In the natural-orbital gauge, the population of eigenstate $k$ on the qubit state $\vec n=(n_0,n_1,n_2,n_3)\in\{0,1\}^4$ can be read directly from the configuration-interaction coefficients: $P^{(k)}_{\vec n} \equiv |C^{(k)}_{\vec n}|^2$.

The five lowest eigenstates are identified directly with $\ket{0000}$ being the ground state and the four qubit states with a single excitation being the first four excited states, in the energy order guaranteed by the constraints $\omega^0<\omega^1<\omega^2<\omega^3$. The remaining eleven qubit states do not have a fixed placement among the eigenstates, since higher excited states start to intrude. Each of these eleven states is instead paired with one of the remaining 45 eigenstates by a weighted linear-sum (Hungarian) assignment, maximizing the total population summed over qubit states with weights $w_{\vec n}=1$ for the six doubles and $w_{\vec n}=0.3$ for the triples and the quadruple $\ket{1111}$. Writing $k_{\vec n}$ for the eigenstate corresponding to qubit state $\vec n$, we define the idle-point cost function as
\begin{equation}
  L_{\mathrm{idle}}(\boldsymbol{\lambda}) = \sum_{\vec n} w_{\vec n}\bigl[1 - P^{(k_{\vec n})}_{\vec n}\bigr],
  \label{eq:loss_idle}
\end{equation}
with $w_{\vec n}=3$ for $\ket{0000}$ and the singles, $1$ for the six doubles, and $0.3$ for the triples and $\ket{1111}$.

In the optimization for a target operating point, the two eigenstates corresponding to the active pair $(i,j)$ are instead required to be maximally mixed. We write $k_i\equiv k_{\vec n_i}$ and $k_j\equiv k_{\vec n_j}$ for the two corresponding eigenstates. The cost function for the target operating point is then
\begin{equation}
  L_{\mathrm{target}} = L_{\mathrm{active}} + L_{\mathrm{spectator}},
  \label{eq:loss_target}
\end{equation}
where $L_{\mathrm{spectator}}$ has the form of Eq.~\eqref{eq:loss_idle}, summed only over the 14 qubit states $\vec n\notin\{\vec n_i,\vec n_j\}$, with weights $w_{\vec n}=3$ for $\ket{0000}$, $1$ for the singles, $0.5$ for the doubles, and $0.15$ for the triples and $\ket{1111}$. The doubles, triples and $\ket{1111}$ are assigned to eigenstates as at the idle point. $L_{\mathrm{active}}$ is a term that penalizes deviations from maximal mixing for the active pair:
\begin{equation}
L_{\mathrm{active}} = W\Bigl[\bigl(P_{i\to i}-\tfrac12\bigr)^2 + \bigl(P_{i\to j}-\tfrac12\bigr)^2 + \bigl(P_{j\to i}-\tfrac12\bigr)^2 + \bigl(P_{j\to j}-\tfrac12\bigr)^2\Bigr],
\end{equation}
with $W=5$ and $P_{i\to j}\equiv P^{(k_i)}_{\vec n_j}$, i.e. the absolute squared overlap between the eigenstate $k_i$ and the qubit state $\ket{\vec{n}_j}$. Only the two active wells are optimized, while the two spectator wells are held at their idle values. For nearest-neighbor pairs the frequencies of the active wells are additionally kept at least $\SI{2}{\giga\hertz}$ from those of the spectators. This gap constraint is relaxed for the long-range pair $(0,3)$, since the frequencies of the active wells must cross those of the frozen middle wells to reach resonance.

Both cost functions are minimized with the derivative-free trust-region solver Py-BOBYQA~\cite{cartisImprovingFlexibilityRobustness2019, cartisEscapingLocalMinima2022} with global restarts enabled. The frequencies and anharmonicities are restricted to the ranges $\omega/2\pi\in[\num{5}, \num{15}]~\si{\giga\hertz}$ and $\beta/2\pi\in[\num{0.05}, \num{2}]~\si{\giga\hertz}$ of Sec.~\ref{sec:op-points}. At the idle point, where all eight parameters are optimized, we do not optimize the frequencies directly but use a stick-breaking reparameterization. We optimize a set of fractions $t_\sigma\in[0,1]$, one per well, and build the frequencies up one well at a time. $\omega^0$ is placed within the allowed range according to $t_0$, and each subsequent $\omega^\sigma$ is placed above $\omega^{\sigma-1}$, within whatever range remains, according to $t_\sigma$. Every choice of fractions therefore yields frequencies satisfying $\omega^0<\omega^1<\omega^2<\omega^3$, so this ordering holds by construction rather than being enforced as a penalty. At the target points, the four parameters of the active wells are optimized directly, and the gap to the spectators is imposed through the bounds on the active frequencies. Each cost evaluation is a full imaginary-time run, warm-started from the converged states of the previous evaluation.

\subsubsection{Optimized operating points}

Here we present the results of the operating-point optimization with one electron per well. Table~\ref{tab:idle_point} lists the frequencies and anharmonicities of the optimized idle operating point and the value of $L_{\mathrm{idle}}$ (Eq.~\eqref{eq:loss_idle}) it reaches. The lowest population attained by any of the 16 qubit states is $0.994$, close to the ideal value of $1$. Table~\ref{tab:target_points} lists the corresponding active-well parameters for each target operating point, the value of $L_{\mathrm{target}}$ (Eq.~\eqref{eq:loss_target}), and the populations of the active pair, which approach the ideal value of $1/2$ at perfect mixing. 

\begin{table}[h]
\caption{Optimized idle-point well parameters, reaching $L_{\mathrm{idle}}=0.030$ with a minimum population of $0.994$ over all 16 qubit states.}
\label{tab:idle_point}
\begin{ruledtabular}
\begin{tabular}{lcccc}
Well & 0 & 1 & 2 & 3 \\
\hline
$\omega^\sigma/2\pi$ (GHz) & 7.69 & 10.67 & 13.79 & 14.96 \\
$\beta^\sigma/2\pi$ (GHz) & 0.295 & 0.563 & 0.562 & 1.971 \\
\end{tabular}
\end{ruledtabular}
\end{table}

\begin{table}[h]
\caption{Optimized active-well parameters at each target operating point. The spectator wells are held at their idle values.}
\label{tab:target_points}
\begin{ruledtabular}
\begin{tabular}{lcccc}
Active pair $(i,j)$ & $(0,1)$ & $(1,2)$ & $(2,3)$ & $(0,3)$ \\
\hline
$\omega^i/2\pi$ (GHz) & 11.75 & 12.96 & 13.97 & 7.92 \\
$\omega^j/2\pi$ (GHz) & 11.75 & 12.96 & 14.37 & 8.23 \\
$\beta^i/2\pi$ (GHz) & 1.323 & 1.977 & 1.921 & 0.263 \\
$\beta^j/2\pi$ (GHz) & 1.538 & 2.000 & 1.254 & 0.071 \\
$L_{\mathrm{target}}$ & 1.061 & 1.070 & 1.083 & 1.016 \\
$P_{i\to i}$ & 0.553 & 0.561 & 0.438 & 0.546 \\
$P_{i\to j}$ & 0.446 & 0.438 & 0.562 & 0.453 \\
$P_{j\to i}$ & 0.447 & 0.439 & 0.562 & 0.453 \\
$P_{j\to j}$ & 0.552 & 0.561 & 0.438 & 0.546 \\
\end{tabular}
\end{ruledtabular}
\end{table}

The values of $L_{\mathrm{target}}$ are close to one although the active pairs are well mixed. They are dominated by the six qubit states in which exactly one active well and at least one spectator are excited, e.g.\ $\ket{1010}$ and $\ket{0110}$ for the pair $(0,1)$. These states are resonant in pairs as well and mix as the gate requires, keeping $0.55$--$0.63$ of their population. As $L_{\mathrm{spectator}}$ counts them as spectator states, they contribute $0.95$--$0.99$, while $L_{\mathrm{active}}$ contributes $0.04$--$0.08$. At the target point of the pair $(1,2)$, the optimum lies on the constraints: both active frequencies are exactly $\SI{2}{\giga\hertz}$ below that of well~3, and $\beta^j$ is at its upper bound of $\SI{2}{\giga\hertz}$. This point is therefore set by the constraints rather than by a minimum of the cost function. The other target points lie inside the allowed region.

\subsection{Three electrons per well}
\label{app:op-points-3e}
The procedure with three electrons per well follows the one-electron case, with the multilayer ansatz of Sec.~\ref{sec:ml} in place of the single-layer one, using 14 orbitals and 4 SBS in every well.

\subsubsection{Constructing the eigenstates}
\label{app:3e-construct}
With three electrons per well, the excitation spectrum of each well is much denser, and the higher excited states intrude already among the single excitations. The states carrying the state-averaging weights of Sec.~\ref{sec:stateavg} can therefore not be identified by their energy order as with one electron per well. At each ITE step we instead match them to the eigenvectors of the top-layer Hamiltonian (Eq.~\eqref{eq:ml-A}) by their populations on the qubit states (Sec.~\ref{app:3e-locate}).

We fix the gauge between the SBS and the top tensor by diagonalizing the state-averaged SBS density matrix $\rho^\sigma$ of Eq.~\eqref{eq:ml-rho} and working in its eigenstates, which we call the natural SBS. The two natural SBS of each well with the largest occupations are taken as its local basis states $\ket{0}_\sigma$ and $\ket{1}_\sigma$, which label the qubit states $\ket{n_0n_1n_2n_3}$. This definition is unambiguous only if the two occupations differ, since otherwise any rotation between the two natural SBS is an equally valid eigenbasis of $\rho^\sigma$. The occupations are controlled by the state-averaging weights $w_N$. In the idle-point optimization, the relaxed states and their weights are the same as with one electron per well, i.e.\ $\ket{0000}$ with weight $1/3$ and the four single excitations with weight $1/6$ each. To see how the weights set the occupations, consider relaxed eigenstates that are products of these local basis states, as is approximately the case for weakly hybridized qubit states. Then $\rho^\sigma$ is diagonal, and the occupation of level $j$ is the total weight of the states in which well $\sigma$ is in level $j$. For example, tracing out wells 1--3 at the idle point gives $\ket{1}_0\bra{1}_0$ for $\ket{1000}$ and $\ket{0}_0\bra{0}_0$ for each of the other four states, so
\begin{equation}
  \rho^0 = \tfrac56\,\ket{0}_0\bra{0}_0 + \tfrac16\,\ket{1}_0\bra{1}_0 ,
\end{equation}
and the two occupations differ. The same holds for the other wells.

Near a resonance, however, the orbitals and SBS optimized for $\ket{0000}$ and the four single excitations describe the other qubit states poorly. As all 16 qubit states enter $L_{\mathrm{target}}$, we relax all of them for the target operating points. Each well is then excited in half of the states, so equal weights would make the two occupations equal. We instead use weights proportional to $2^{-n}$, where $n$ is the number of excited wells, which gives $\ket{0}_\sigma$ twice the occupation of $\ket{1}_\sigma$. At the idle point, relaxing all 16 states from the five states of the optimization changes $L_{\mathrm{idle}}$ by only $0.25\%$, so the idle optimization was not repeated. Relaxing the 16 states from the uncoupled product states instead gives a slightly lower state-averaged energy and lowers $L_{\mathrm{idle}}$ from $0.0101$ to $0.0090$. Following the rule of Sec.~\ref{app:3e-locate}, these are the idle states used below.

\subsubsection{Locating operating points}
\label{app:3e-locate}
Analogous to the 1 electron case, the population of eigenstate $k$ on the qubit state $\vec n$ is read from the top tensor $A^{(k)}$ (Sec.~\ref{sec:ml}) in the natural-SBS gauge. It is given by $P^{(k)}_{\vec n}=\bigl|A^{(k)}_{J}\bigr|^2$ with $J=\vec n$, i.e.\ the product of the local basis states $\ket{n_\sigma}_\sigma$. However, the single excitations are no longer the lowest states after $\ket{0000}$. All 16 qubit states are therefore assigned to eigenstates by the weighted linear-sum assignment, not only the eleven higher ones. The cost functions, Eqs.~\eqref{eq:loss_idle} and~\eqref{eq:loss_target}, and the optimizer are otherwise unchanged.

The target optimizations start from the one-electron target points of Table~\ref{tab:target_points}, with the spectator wells at their three-electron idle values. At these points the active pair is off resonance. To start the optimization from a mixed pair, we first bring the pair into resonance by a one-dimensional root search along $\beta^j$ for the zero of the energy gap between the two eigenstates assigned to the active pair. The search ends when the pair is mixed, $P_{i\to i}\le0.75$, and the optimization starts from that point.

Two further safeguards are needed. First, $L_{\mathrm{active}}$ is minimized by any eigenstate with $P_{i\to i}\approx\tfrac12$, whichever state carries the remaining population. It can therefore be lowered by an active qubit state hybridizing with an intruding state instead of with its partner. We reject evaluations in which either of the two eigenstates keeps less than $0.95$ of its population on the active pair, $P_{i\to i}+P_{i\to j}<0.95$, by assigning them a large cost. Second, where no warm start is available, the states are relaxed both from the uncoupled product states of Sec.~\ref{app:1e-construct} and from the idle-point states, and the result with the lower state-averaged energy is kept.

\subsubsection{Optimized operating points}
Table~\ref{tab:idle_point_3e} lists the optimized idle operating point with three electrons per well. It reaches $L_{\mathrm{idle}}=0.0089$, with a lowest population of $0.9985$, on $\ket{1111}$, compared with $0.030$ and $0.994$ with one electron per well. Table~\ref{tab:target_points_3e} lists the target operating points of the three nearest-neighbor pairs. The active pairs mix 54/46 to 55/45, as with one electron per well. Again, the active frequencies of the pair $(1,2)$ lie exactly $\SI{2}{\giga\hertz}$ below that of well~3, and here this also holds for $\omega^j$ of the pair $(0,1)$, which lies $\SI{2}{\giga\hertz}$ below $\omega^2$. The values of $L_{\mathrm{target}}$ exceed those with one electron per well by $0.22$--$0.31$, mainly because the qubit states with both active wells excited keep only $0.68$--$0.77$ of their population. These states hybridize with nearby states in which one active well is excited beyond its qubit levels (Sec.~\ref{sec:threewell}). The parameters in both tables were optimized with 64 grid points per well. The calculations with three electrons per well otherwise use 80 points (App.~\ref{app:grid}), and the cost functions and populations in the tables are evaluated on this finer grid, without reoptimizing the parameters.

\begin{table}[h]
\caption{Optimized idle-point well parameters with three electrons per well, reaching $L_{\mathrm{idle}}=0.0089$ with a minimum population of $0.9985$ over all 16 qubit states.}
\label{tab:idle_point_3e}
\begin{ruledtabular}
\begin{tabular}{lcccc}
Well & 0 & 1 & 2 & 3 \\
\hline
$\omega^\sigma/2\pi$ (GHz) & 7.49 & 10.52 & 13.78 & 14.96 \\
$\beta^\sigma/2\pi$ (GHz) & 0.282 & 0.560 & 0.673 & 1.959 \\
\end{tabular}
\end{ruledtabular}
\end{table}

\begin{table}[h]
\caption{Optimized active-well parameters at each target operating point with three electrons per well. The spectator wells are held at their idle values (Table~\ref{tab:idle_point_3e}).}
\label{tab:target_points_3e}
\begin{ruledtabular}
\begin{tabular}{lccc}
Active pair $(i,j)$ & $(0,1)$ & $(1,2)$ & $(2,3)$ \\
\hline
$\omega^i/2\pi$ (GHz) & 11.75 & 12.96 & 13.97 \\
$\omega^j/2\pi$ (GHz) & 11.78 & 12.96 & 14.37 \\
$\beta^i/2\pi$ (GHz) & 1.323 & 1.975 & 1.921 \\
$\beta^j/2\pi$ (GHz) & 1.307 & 1.978 & 1.805 \\
$L_{\mathrm{target}}$ & 1.374 & 1.308 & 1.301 \\
$P_{i\to i}$ & 0.541 & 0.554 & 0.553 \\
$P_{i\to j}$ & 0.458 & 0.446 & 0.447 \\
$P_{j\to i}$ & 0.459 & 0.446 & 0.447 \\
$P_{j\to j}$ & 0.541 & 0.554 & 0.553 \\
\end{tabular}
\end{ruledtabular}
\end{table}

\section{Single-qubit gates}
\label{app:single-qubit}
The routed gates of Sec.~\ref{sec:longrange-test} combine propagated \sqiswap{}
gates with single-qubit gates that are not propagated. Instead, each
single-qubit gate is applied as an instantaneous transformation of the
wavefunction parameters of the well it acts on. In this section we describe this transformation,
explain why it is not exact, and quantify the error it adds to the routed gates. Finally, Sec.~\ref{app:rz-angles} describes how the angles of the $R_z$ rotations in Eq.~\eqref{eq:fidelity} are chosen. These rotations accompany every two-qubit gate, including the gates applied on their own in Sec.~\ref{sec:mf-res}, although the choice of angles only affects the fidelity of a circuit.

\subsection{Implementation}
Each well $\sigma$ is assigned two orthonormal \emph{qubit vectors}
$\ket{q^{\sigma}_0}$ and $\ket{q^{\sigma}_1}$ in the $N_\sigma$-electron space
of the well. A single-qubit gate with the $2\times2$ matrix $u$ acting on
qubit $\sigma$ is applied as the operator
\begin{equation}
  \hat U^{\sigma}(u) = 1
  - \sum_{a=0}^{1}\ket{q^{\sigma}_a}\bra{q^{\sigma}_a}
  + \sum_{a,b=0}^{1} u_{ab}\,\ket{q^{\sigma}_a}\bra{q^{\sigma}_b},
  \label{eq:sq-layer}
\end{equation}
which rotates the qubit vectors by $u$ and leaves the rest of the space of
well $\sigma$ unchanged. The operator acts on the parameters of the ansatz
that span this space, and the configuration coefficients and the other wells
are left unchanged.

\emph{One electron per well.} The qubit vectors are single-particle orbitals
$q^{\sigma}_a$, and the operator acts on the orbital coefficients of
Eq.~\eqref{eq:bexp},
$b^{\sigma}_{kn}\to\sum_{k'}U^{\sigma}_{kk'}\,b^{\sigma}_{k'n}$ with
$U^{\sigma}_{kk'}=\langle\chi^{\sigma}_k|\hat U^{\sigma}(u)|\chi^{\sigma}_{k'}\rangle$.
The orbitals are taken from the idle eigenstates. $q^{\sigma}_0$ is the
dominant natural orbital of well $\sigma$ in the eigenstate labeled
$\ket{0000}$, and $q^{\sigma}_1$ the dominant natural orbital in the
eigenstate where only qubit $\sigma$ is excited. In both states the other
wells are in their ground state, so the two orbitals are the least affected by
the neighboring wells. They are orthogonal to within
$|\langle q^{\sigma}_0|q^{\sigma}_1\rangle|\le\num{2.4e-5}$ and are
orthonormalized before use.

\emph{Three electrons per well.} The routed circuits are simulated only with one electron per well (Sec.~\ref{sec:longrange-test}). We nevertheless describe how the construction carries over to three electrons per well, where a rotation of single-particle orbitals is not sufficient. The qubit state $\ket{1}_\sigma$ is the center-of-mass excitation of the well (Sec.~\ref{sec:threewell}), a
superposition of several determinants whose orbitals differ from those of
$\ket{0}_\sigma$. No rotation of a single orbital pair reaches it: for the
Hadamard gate, the one-body operator gives an average gate fidelity of only
$0.42$--$0.47$ on the 16 qubit states, against $0.998$--$0.9998$ for
the many-body operator below. We therefore take the qubit vectors to be
many-body states in the determinant space of the well, $\ket{q^{\sigma}_a}=\sum_{I_\sigma}q^{\sigma}_{I_\sigma a}\ket{I_\sigma}$, and apply
Eq.~\eqref{eq:sq-layer} to the SBS coefficients of Eq.~\eqref{eq:sbs},
$B^{\sigma}\to U^{\sigma}B^{\sigma}$, with $U^{\sigma}$ the matrix of
$\hat U^{\sigma}(u)$ in the determinant basis. The orbitals and the top tensor $A$ are left
unchanged. The vectors $\ket{q^{\sigma}_a}$ are the dominant natural SBS of
well $\sigma$ in the same two idle eigenstates as above, i.e.\ the local basis states of
Sec.~\ref{app:3e-construct}. Since the orbitals change during the propagation, the vectors, defined at $t=0$, are first
projected onto the determinant space of the current orbitals and then
orthonormalized (Löwdin). For $N_\sigma=1$ this construction is the one above.

In both cases, the orbitals and, at three electrons per well, the SBS are shared by all
states in the state-averaged propagation (Sec.~\ref{sec:stateavg}), so a single update applies the
gate to all 16 propagated basis states at once.

Equation~\eqref{eq:sq-layer} is unitary for any unitary $u$, so the norm of the
propagated states is conserved. Because the qubit vectors are orthonormal,
the operators also compose exactly,
$\hat U^{\sigma}(u)\,\hat U^{\sigma}(v)=\hat U^{\sigma}(uv)$, and consecutive
single-qubit gates on the same well can be merged into a single transformation without approximation. For instance in the SWAP blocks of Fig.~\ref{fig:swap_decomposition}, the implicit $R_z$ rotations of the \sqiswap{} gate can be merged with the $H$ and $S^{\dagger}$ gates, so that the serial and parallel routings contain 14 and 8 orbital transformations, respectively.

\subsection{Error}
Equation~\eqref{eq:sq-layer} is an exact single-qubit gate only if every
eigenstate at the idle operating point separates into a product of
single-well states. Because of the inter-species interaction, the eigenstates separate only
approximately. The reduced density matrix of each well has some weight
outside the span of $\ket{q^{\sigma}_0}$ and $\ket{q^{\sigma}_1}$, and its dominant
natural orbitals (SBS) vary slightly with the state of the other wells. The part outside the span is not rotated by
the gate, so it leaves the computational subspace as leakage and contributes a
phase error.

\emph{One electron per well.} To quantify the accumulated error, we evaluate the $16\times16$ product model
of Fig.~\ref{fig:methods_fid} with ideal single-qubit gates and with those of Eq.~\eqref{eq:sq-layer}.
The latter lower $\bar F_{16}$ from $0.931$ to $0.875$ for the serial routing, from $0.891$ to $0.828$ for the parallel routing, and from $0.915$ to $0.836$ for the synchronized routing.

\emph{Three electrons per well.} We evaluate the construction on the 16 idle eigenstates of the operating-point optimization (Sec.~\ref{app:op-points-3e}, 64 grid points per well). The dominant natural SBS carry occupations of $0.999987$--$0.999999$ for $\ket{0}_\sigma$ and $0.99940$--$0.99994$ for $\ket{1}_\sigma$, and they vary by at most $1.8\times10^{-4}$ over the 16 states. Applied to these states, the Hadamard gate reaches $\bar F=0.998$--$0.9998$ depending on the well, and the $HS^\dagger$ block on a pair of neighboring wells $0.9986$--$0.9993$.

\subsection{Angles of the $R_z$ rotations}
\label{app:rz-angles}
The $R_z$ rotations before and after each \sqiswap{} gate are the maximizing angles of Eq.~\eqref{eq:fidelity} for the gate propagated on its own. This maximum is degenerate. The equal rotation $R_z^{\otimes2}(\varphi)=R_z(\varphi)\otimes R_z(\varphi)$ of both wells multiplies $\ket{00}$, $\ket{01}$, $\ket{10}$ and $\ket{11}$ by $e^{-i\varphi}$, $1$, $1$ and $e^{i\varphi}$. The ideal gate $U$ mixes only $\ket{01}$ and $\ket{10}$, which acquire the same phase, and therefore commutes with $R_z^{\otimes2}(\varphi)$. For any $A$, and in particular for $A=[R_z(\theta_a)\otimes R_z(\theta_b)]\,M\,[R_z(\phi_a)\otimes R_z(\phi_b)]$ with the propagated gate matrix $M$ of Eq.~\eqref{eq:fidelity}, we therefore have that
\begin{equation}
  \mathrm{Tr}\bigl(U^\dagger R_z^{\otimes2}(\varphi)\,A\,R_z^{\otimes2}(-\varphi)\bigr)
  =\mathrm{Tr}\bigl([R_z^{\otimes2}(-\varphi)\,U\,R_z^{\otimes2}(\varphi)]^\dagger A\bigr)
  =\mathrm{Tr}\bigl(U^\dagger A\bigr).
  \label{eq:rz-degeneracy}
\end{equation}
A common angle $\varphi$ can thus be moved from the rotations before the gate to those after it without changing $\bar F_4$. The choice of $\varphi$ nevertheless matters for the parts of the propagated state that the ideal gate lacks. Transitions between states that acquire different phases, such as from $\ket{00}$ to $\ket{01}$, change phase with $\varphi$. Since the rotations act only on the qubit states, the phase of the population that has left the qubit subspace also changes with $\varphi$, relative to the qubit states. In a circuit, this population can return in a later gate and interfere with the population that remained. The fidelity of a circuit therefore depends on $\varphi$, although that of each gate does not.

We fix $\varphi$ by choosing equal rotations before and after each gate. Since the control schedule is symmetric in time and the Hamiltonian and the qubit states are real, the exact gate matrix is symmetric, and so is the ideal gate $U$. The propagated $M$ is symmetric to within the truncation error. Transposition therefore maps a maximizing set of angles to another one in which the rotations before and after the gate are exchanged. As the maximum is unique up to the shift by $\varphi$, the two sets differ only by such a shift, so a maximum with equal rotations before and after the gate exists. This condition fixes $\varphi$ up to $\pi$, and we take the solution with the smaller $\theta_a^2+\theta_b^2$, with the angles taken in $(-\pi,\pi]$. When two gates run concurrently and one finishes early, the rotations after that gate are determined for the gate followed by the free evolution of its pair over the remaining time.

\section{Timing of the gate schedule}
\label{app:gate-timing}
The control schedule of Eq.~\eqref{eq:time_dep_schedule} has two timing parameters, $t_{\mathrm{ramp}}$ and $t_{\mathrm{hold}}$. This section derives the window in which each is swept in Sec.~\ref{sec:mf-res} and lists the resulting control schedules. The derivation uses only the eigenstates of the idle and target operating points, without fitting and without any real-time propagation. Times are given in ns, and the dimensionless frequency $\omega$ below is $2\pi$ times the frequency in GHz.

\subsection{Hold time}
\label{app:hold-time}
At the target operating point of a gate between wells $a$ and $b$, two eigenstates are the superpositions $(\ket{1_a}\pm\ket{1_b})/\sqrt2$ of the single-excitation qubit states, separated by the splitting $\Delta$. In the subspace $\{\ket{01},\ket{10}\}$ of the active pair the Hamiltonian is thus $H=(\Delta/2)X$ up to a constant, and a hold of duration $T$ rotates the pair by $\theta=\Delta \cdot T$. The gate is the identity on $\ket{00}$ and $\ket{11}$ and, in the exchange block,
\begin{equation}
  U(\theta)=\cos\tfrac{\theta}{2}\,\openone-i\sin\tfrac{\theta}{2}\,X .
\end{equation}
The \sqiswap{} gate corresponds to $\theta=\pi/2$, which gives the hold-time center $T_0=\pi/(2\Delta)$.

The width of the hold-time window is set by how quickly the fidelity drops when the rotation angle misses $\pi/2$ by $\delta\theta$. For a leakage-free gate with this error, $M=U(\pi/2+\delta\theta)$, so that $\mathrm{Tr}(MM^\dagger)=4$ and $U(\pi/2)^\dagger M=U(\delta\theta)$. The maximum in Eq.~\eqref{eq:fidelity} is then reached without $R_z$ rotations, and
\begin{equation}
  \bar F_4(\delta\theta)=\frac{4+\bigl(2+2\cos(\delta\theta/2)\bigr)^2}{20}\simeq 1-\frac{\delta\theta^2}{10},
\end{equation}
so that an infidelity budget $\varepsilon$ admits $\delta\theta=\sqrt{10\varepsilon}$, i.e.\ a hold-time width $\delta T=\sqrt{10\varepsilon}/\Delta$. Leakage that does not depend on the hold time lowers $\bar F_4$ but leaves this width almost unchanged. Errors that grow with the hold time, such as accumulated leakage and phase errors caused by the spectators, instead make the computed optima narrower than $\delta T$ (Sec.~\ref{app:sweep-results}). We therefore use $\delta T$ for $\varepsilon=10^{-2}$ to set the range of the sweep, $\pm2\,\delta T$ around the center, and sample this range in steps of $\delta T/3$.

\subsection{Ramps}
The coupling is only partly on during the ramps, so the rotation accumulated over a gate is not $\Delta\cdot t_{\mathrm{hold}}$ but $\theta=\Delta\cdot T_{\mathrm{eff}}$, with the effective hold time $T_{\mathrm{eff}}=t_{\mathrm{hold}}+\alpha\,t_{\mathrm{ramp}}$. The ramp of Eq.~\eqref{eq:ramp} depends on time only through $u=t/t_{\mathrm{ramp}}$, so the rising edges of the schedule for different ramp times differ only by a rescaling of the time axis and are all given by $f(u)=\tfrac12\bigl[1+\mathrm{erf}\bigl(6(u-\tfrac12)\bigr)\bigr]$. Each ramp therefore contributes an amount proportional to $t_{\mathrm{ramp}}$, and $\alpha$ is independent of it. If the splitting during the ramps is taken to be proportional to the edge, $\Delta(t)=\Delta\cdot f(t/t_{\mathrm{ramp}})$, the two ramps add $2\Delta\cdot t_{\mathrm{ramp}}\int_0^1 f(u)\,\mathrm{d}u$ to the rotation. Since $f(u)-\tfrac12$ is antisymmetric about $u=\tfrac12$, the integral is $\tfrac12$ and $\alpha=1$, i.e.\ each ramp counts as half its duration at full coupling. We use this value to center the sweep at $t_{\mathrm{hold}}=T_0-t_{\mathrm{ramp}}$. It is a rule for centering and not a prediction of the optimum: at the computed optima, $T_{\mathrm{eff}}$ lies within about one $\delta T$ of $T_0$ (Tables~\ref{tab:timing} and~\ref{tab:schedules}).

The ramp time is bounded on both sides. It must be long enough not to excite the wells, i.e.\ slow compared with their transition frequencies. The smallest of these is the one-body $0\to1$ spacing $\omega+\beta$ of the active wells (Sec.~\ref{app:anharm}), which for all four gates is smallest at the idle point. We denote its value there by $\omega_{\min}$ and take $t_{\mathrm{ramp}}\ge 10/\omega_{\min}$, where the prefactor is chosen rather than derived. The ramp time must also be short enough that the two ramps alone do not complete the rotation. With $\alpha=1$, the centered hold time $T_0-t_{\mathrm{ramp}}$ is negative for $t_{\mathrm{ramp}}>T_0=\pi/(2\Delta)$. We take $t_{\mathrm{ramp}}\le1/\Delta$, which keeps it positive with some margin. We sample the window between $10/\omega_{\min}$ and $1/\Delta$ geometrically with six values, since near the lower edge the diabatic excitation decreases rapidly with $\omega_{\min}t_{\mathrm{ramp}}$. For \sqiswap{03}, the frequency of well~3 passes those of wells~1 and~2 during the ramps, which the lower bound does not account for.

\subsection{Resulting grids}
Table~\ref{tab:timing} lists the resulting windows. The sweep is performed in $(t_{\mathrm{ramp}},T_{\mathrm{eff}})$ with 13 values of $T_{\mathrm{eff}}$ in steps of $\delta T/3$ per ramp time, with all times rounded to the \SI{1.59}{\pico\second} checkpoint grid of the propagation. This gives 77 points for each nearest-neighbor gate, after removing the one point with $t_{\mathrm{hold}}<0$. For the \sqiswap{03} gate, the window of ramp times is very wide, and only the first three ramp values are swept in order to limit the cost (39 points). For the best ramp times, we refine the hold time around the maximum on this grid with four additional points, spaced by the step $\delta T/(3\sqrt{10})$ that corresponds to the infidelity budget $\varepsilon=10^{-3}$ of Sec.~\ref{app:hold-time}.

\begin{table}
\caption{Timing windows of the sweeps in Sec.~\ref{sec:mf-res}. $\Delta/2\pi$ is the splitting at the target operating point, $T_0=\pi/(2\Delta)$ the hold-time center, and $\delta T$ the width for $\varepsilon=10^{-2}$. The sweep covers $T_0\pm2\,\delta T$ in steps of $\delta T/3$. The ramp times span $[10/\omega_{\min},1/\Delta]$.}
\label{tab:timing}
\begin{ruledtabular}
\begin{tabular}{lcccc}
Gate & $\Delta/2\pi$ (MHz) & $T_0$ (ns) & $\delta T$ (ns) & $t_{\mathrm{ramp}}$ range (ns) \\
\hline
\sqiswap{01} & 121.5 & 2.06 & 0.41 & 0.20--1.31 \\
\sqiswap{12} & 109.9 & 2.27 & 0.46 & 0.14--1.45 \\
\sqiswap{23} & 101.9 & 2.45 & 0.49 & 0.11--1.56 \\
\sqiswap{03} & 7.5 & 33.4 & 6.7 & 0.20--21.3 \\
\end{tabular}
\end{ruledtabular}
\end{table}

\subsection{Evaluation of the sweeps}
To reduce the cost of the sweeps of Sec.~\ref{sec:mf-res}, we do not propagate each point from $t=0$. For each ramp time, a single propagation ramps up to the target point and stays there for the longest hold time of the sweep, storing the state every \SI{1.59}{\pico\second}. The propagation for a given hold time is then restarted from the stored state at $t=t_{\mathrm{gate}}-2t_{\mathrm{ramp}}$, one ramp time before its ramp-down begins, and then follows the rest of its control schedule. Up to this time, the ramp-down edge, centered at $t_{\mathrm{gate}}-\tfrac12 t_{\mathrm{ramp}}$, contributes less than $\mathrm{erfc}(9)\approx4\times10^{-37}$, so the restarted propagation agrees with a propagation of the full schedule to machine precision. Each additional hold time thus costs only a propagation over $2t_{\mathrm{ramp}}$.

\subsection{Results of the sweeps}
\label{app:sweep-results}
\begin{figure*}[h]
\centering
\includegraphics{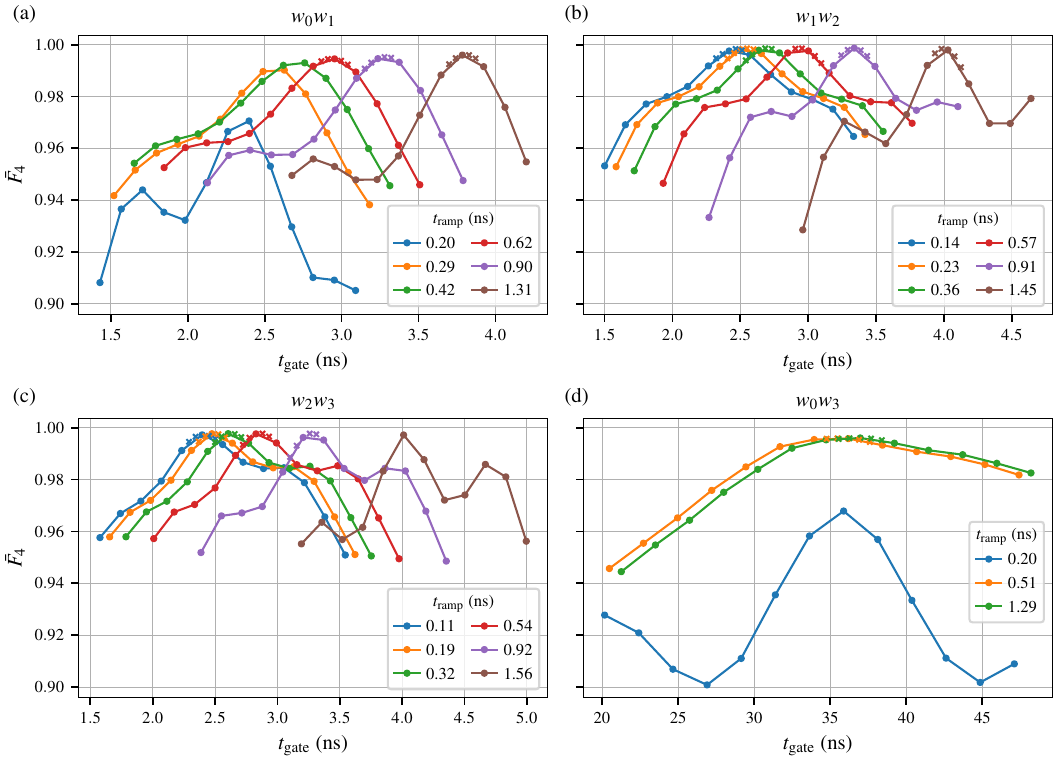}
\caption{Average gate fidelity $\bar F_4$ against gate duration $t_{\mathrm{gate}}=2t_{\mathrm{ramp}}+t_{\mathrm{hold}}$ in the mean-field sweeps of (a)~\sqiswap{01}, (b)~\sqiswap{12}, (c)~\sqiswap{23} and (d)~the direct \sqiswap{03} gate, with one orbital in each spectator well and four in each active well. Colors denote the ramp time $t_{\mathrm{ramp}}$. Circles connected by lines mark the coarse grid of Table~\ref{tab:timing}, and crosses the finer hold-time sweeps around the maximum for selected ramp times.}
\label{fig:1e_mf_sweep}
\end{figure*}
Figure~\ref{fig:1e_mf_sweep} shows the mean-field sweep over the ramp and hold times for all four gates. For each ramp time, the fidelity has one dominant maximum as a function of the hold time. The maxima of different ramp times lie at nearly the same effective hold time $T_{\mathrm{eff}}=t_{\mathrm{hold}}+t_{\mathrm{ramp}}$, within \SI{0.3}{\nano\second} for the nearest-neighbor gates and \SI{0.7}{\nano\second} for \sqiswap{03}. In terms of the gate duration, they therefore shift by about the ramp time. The coarse sweep cannot resolve which ramp time gives the best fidelity, since moving the hold time one step, $\delta T/3$, away from the maximum lowers $\bar F_4$ by $2\times10^{-3}$ to $4\times10^{-3}$, more than the maxima of the best ramp times differ. We therefore rank the ramp times by the finer hold-time sweeps (crosses).

The refined sweeps show that the ramp time matters little for most gates. For \sqiswap{12} and \sqiswap{23}, the maxima of the three best ramp times agree within $1\times10^{-4}$, and the shortest ramp times (\SI{0.14}{\nano\second} and \SI{0.11}{\nano\second}) lose only $3.4\times10^{-4}$ and $5.0\times10^{-4}$ against the best one. These gates can therefore be made faster at little cost in fidelity. We nevertheless use the optimum of each sweep in Fig.~\ref{fig:mf} and Table~\ref{tab:schedules}. For \sqiswap{01}, the fidelity increases with the ramp time up to the longest swept value of \SI{1.31}{\nano\second}, with a difference of $1.5\times10^{-3}$ between the three best ramp times, and the shortest ramp time (\SI{0.20}{\nano\second}) reaches only $\bar F_4=0.971$. The optimum for this gate is thus at the edge of the swept window, and a longer ramp could improve it slightly. For the direct \sqiswap{03} gate, the shortest of the three swept ramp times reaches only $0.968$, while the other two differ by $1.5\times10^{-4}$, which is far below the mean-field error of this gate (Sec.~\ref{sec:mf-res}).

\subsection{Final control schedules}
Table~\ref{tab:schedules} lists the ramp and hold times of all gates. With one electron per well, the individually optimized schedules are the maxima of the sweeps in Fig.~\ref{fig:1e_mf_sweep}, and the synchronized schedule is the common schedule of the \sqiswap{01} and \sqiswap{23} gates in the synchronized routing (Sec.~\ref{sec:longrange-test}).

With three electrons per well, the sweep is run on the 64-point grid of the operating-point optimization (Sec.~\ref{app:op-points-3e}), with 14 orbitals and 4 SBS per active well and mean-field spectators, and $\omega_{\min}$ is the $1\to2$ spacing, which is here smaller than the $0\to1$ spacing. This sweep does not select a schedule. For every ramp time, $\bar F_4$ is highest at the shortest hold time, where the exchange is only about half completed and $\ket{11}$ retains the most population. We therefore fix the schedule by the rotation angle instead. For each gate, we take the second-shortest ramp time and interpolate the hold time along the sweep to $\theta=\pi/2$. Transferred to the 80-point grid of Sec.~\ref{sec:threewell} by the ratio of the splittings, these schedules give $\theta=\pi/2$ with 22 orbitals per active well to within $5\times10^{-4}$.

\begin{table}[h]
\caption{Control schedules of the gates. $t_{\mathrm{ramp}}$ and $t_{\mathrm{hold}}$ are the ramp and hold times of Eq.~\eqref{eq:time_dep_schedule}, and $t_{\mathrm{gate}}=2t_{\mathrm{ramp}}+t_{\mathrm{hold}}$ is the gate duration, computed before rounding. In the synchronized routing, the \sqiswap{12} gate keeps its individually optimized schedule. With three electrons per well, the hold times are those that give $\theta=\pi/2$ with 22 orbitals per active well.}
\label{tab:schedules}
\begin{ruledtabular}
\begin{tabular}{lccc}
Gate & $t_{\mathrm{ramp}}$ (ns) & $t_{\mathrm{hold}}$ (ns) & $t_{\mathrm{gate}}$ (ns) \\
\hline
\multicolumn{4}{l}{\emph{One electron per well, individually optimized}} \\
\sqiswap{01} & 1.31 & 1.17 & 3.78 \\
\sqiswap{12} & 0.91 & 1.52 & 3.34 \\
\sqiswap{23} & 0.54 & 1.80 & 2.88 \\
\sqiswap{03} & 1.29 & 34.39 & 36.98 \\
\hline
\multicolumn{4}{l}{\emph{One electron per well, synchronized}} \\
\sqiswap{01}, \sqiswap{23} & 0.91 & 1.48 & 3.30 \\
\hline
\multicolumn{4}{l}{\emph{Three electrons per well}} \\
\sqiswap{01} & 0.20 & 1.56 & 1.97 \\
\sqiswap{12} & 0.15 & 2.01 & 2.31 \\
\sqiswap{23} & 0.13 & 2.09 & 2.35 \\
\end{tabular}
\end{ruledtabular}
\end{table}

\section{Convergence of the time integration}
\label{app:propagation-tests}
In this section we describe our verification of the integrator settings of App.~\ref{app:propagation} for the MCTDHF propagations in the main text.

\subsection{One electron per well}
The propagations with one electron per well use the CMF scheme of App.~\ref{app:propagation} in double precision, with $\Delta t=\SI{0.80}{\pico\second}$ for the nearest-neighbor gates and \SI{0.32}{\pico\second} for the direct \sqiswap{03} gate and the routed circuits. For a \SI{2.6}{\nano\second} \sqiswap{12} gate, $\bar F_4$ obtained with \SI{0.80}{\pico\second} differs by less than $3\times10^{-7}$ from that with a ten times smaller step and is thus converged. For the direct gate and some of the routed circuits, however, this step makes the propagation unstable: the occupation of the fourth natural orbital grows until the propagation breaks down. At \SI{0.32}{\pico\second} the same states propagate stably, so the instability is caused by the step and not by the orbital truncation. Reducing the step from 0.32 to \SI{0.16}{\pico\second} changes $\bar F_{16}$ of a test routing by $2\times10^{-4}$. This is an estimate of the error that the step of \SI{0.32}{\pico\second} introduces in the fidelities of the circuits.

\subsection{Three electrons per well}
We verify the integrator settings of App.~\ref{app:propagation-3e} by changing them one at a time in a propagation that covers the ramp of the \sqiswap{12} gate and the first \SI{0.16}{\nano\second} of the hold, with 22 orbitals per active well. The difference between the original and the changed propagation is measured by $1-\bar F_4$ of Eq.~\eqref{eq:fidelity} with $U=1$, applied to the overlap matrix $\braket{\Psi_i|\Psi'_j}$ of their normalized final states. Halving both sub-step lengths changes the result by $1.3\times10^{-6}$, whereas halving $\Delta t$ at fixed sub-step lengths changes it by $5.9\times10^{-9}$. Omitting the Löwdin orthogonalization changes the result by $1.7\times10^{-6}$ and the norms of the final states by up to $8\times10^{-4}$. The longest propagation of Sec.~\ref{sec:threewell} lasts \SI{3.24}{\nano\second}, about ten times longer than the test. If the differences grow quadratically with the propagation time, they remain below $2\times10^{-4}$, which is negligible compared with the infidelities of Sec.~\ref{sec:threewell}. The mixed-precision scheme was tested separately, on a complete \sqiswap{12} gate of \SI{2.58}{\nano\second} with 14 orbitals per active well and 64 grid points, where it agrees with a propagation in double precision to within $2.5\times10^{-9}$.

\renewcommand{\refname}{References}
\bibliography{references}